\documentclass[aps,prd,reprint,secnumarabic,groupedaddress,preprintnumbers,amsmath,amssymb,longbibliography,nofootinbib]{revtex4-1}
\usepackage{latexsym}
\usepackage{graphicx}
\usepackage{verbatim}
\usepackage{amsthm}
\usepackage{float}
\usepackage{amsmath}
\usepackage{booktabs}
\usepackage{graphicx,subfigure}
\usepackage{epstopdf}
\usepackage{indentfirst}
\usepackage{epsfig}
\usepackage{epsfig,subfigure,psfrag}
\usepackage{dcolumn}
\usepackage[T1]{fontenc}
\usepackage{bm}
\usepackage{slashed}
\usepackage{cancel}
\usepackage{color}
\usepackage{tabularx}
\usepackage{hyperref}
\hypersetup{
    colorlinks=true,       
    linkcolor=blue,          
    citecolor=blue,        
    filecolor=blue,      
    urlcolor=blue           
}

\newcommand{\beq}{\begin{equation}}
\newcommand{\eeq}{\end{equation}}
\newcommand{\beqnarray}{\begin{eqnarray}}
\newcommand{\eeqnarray}{\end{eqnarray}}
\newcommand{\OO}{\mathcal{O}}
\newcommand{\lsim}{\lesssim}
\newcommand{\gsim}{\gtrsim}
\newcommand{\chiPT}{\chi\text{PT}}
\newcommand{\RchiT}{\text{R}\chi\text{T}}
\newcommand{\Be}{^8\text{Be}}
\newcommand{\He}{^4\text{He}}
\newcommand{\qu}{q^u_{\text{\tiny PQ}}}
\newcommand{\qd}{q^d_{\text{\tiny PQ}}}
\newcommand{\qe}{q^e_{\text{\tiny PQ}}}
\newcommand{\Br}{\text{Br}}
\newcommand{\ie}{{\it i.e.}}
\newcommand{\thetaS}{\theta_{a\eta_{_{s}}}}
\newcommand{\thetaUD}{\theta_{a\eta_{_{ud}}}}
\newcommand{\thetaPI}{\theta_{a\pi}}
\newcommand{\Tr}{\text{Tr}}

\begin{document}

\title{Signals of the QCD axion with mass of 17~MeV\!$/c^2$:\\Nuclear transitions and light meson decays}

\author{Daniele S.\;M. Alves}
 \email{spier@lanl.gov}
 \affiliation{Theoretical Division, Los Alamos National Laboratory, Los Alamos, NM 87545}

\date{\today}

\begin{abstract}

The QCD axion remains experimentally viable in the mass range of $\OO(10\,$MeV) if (i) it couples predominantly to the first generation of SM fermions; (ii) it decays to $e^+ e^-$ with a short lifetime\footnote{See added note at the end of this article.} $\tau_a\lsim 10^{-13}\;$s; and (iii) it has suppressed isovector couplings, {\it i.e.}, if it is \emph{piophobic}.
Remarkably, these are precisely the properties required to explain recently observed anomalies in nuclear de-excitations, {\it to wit}: the $e^+e^-$ emission spectra of \emph{isoscalar} \emph{magnetic} transitions of~$\Be$ and~$\He$ nuclei showed a ``bump-like'' feature peaked at $m_{e^+e^-}\sim 17$ MeV.  In this article, we argue that on-shell emission of the QCD axion (with the aforementioned properties) provides an extremely well-motivated, compatible explanation for the observed excesses in these nuclear de-excitations.
The absence of anomalous features in other measured transitions is also naturally explained: \emph{piophobic} axion emission is strongly suppressed in isovector magnetic transitions, and forbidden in electric transitions.
This QCD axion hypothesis is further corroborated by an independent observation: a
$\sim2-3\,\sigma$ deviation in the measurement of $\Gamma(\pi^0\to e^+e^-)$ from the Standard Model theoretical expectation. This article also includes detailed estimations of various axionic signatures in rare light meson decays, which take into account contributions from low-lying QCD resonance exchange, and, in the case of rare Kaon decays, the possible effective implementations of $\Delta S=1$ octet enhancement in chiral perturbation theory. These inherent uncertainties of the effective description of the strong interactions at low energies result in large variations in the predictions for hadronic signals of the QCD axion; in spite of this, the estimated ranges for rare meson decay rates obtained here can be probed in the near future in $\eta/\eta^\prime$ and Kaon factories.

\end{abstract}

\preprint{LA-UR-20-27039}
\maketitle

\section{Introduction}
\label{intro}

The past decade has seen a resurgence of interest in the phenomenology of new light particles with feeble interactions with the Standard Model (SM) \cite{Jaeckel:2010ni,Hewett:2012ns,Alexander:2016aln}.
Motivations have been varied, spurred from the growing belief that dark matter might be part of a more complex \emph{dark sector} with additional matter and force carriers \cite{ArkaniHamed:2008qn,Batell:2009yf,Essig:2009nc,Bjorken:2009mm,Batell:2009di}, but also because light dark sectors could be parasitically explored in the broader U.S. and worldwide neutrino program \cite{deNiverville:2011it, Dobrescu:2014ita, Frugiuele:2017zvx, Buonocore:2019esg}.
Experimental signatures of dark sectors
are being searched for by a diverse suite of experiments ranging from beam dumps/fixed targets to meson factories\footnote{See, {\it e.g.}, talks at the kickoff meeting of the RF6 SNOWMASS Working Group, {\it ``Dark Sectors at High Intensities,''} August 12-13, 2020, \url{https://indico.fnal.gov/event/44819/}.}. 
This effort drew on the legacy of an earlier, very active period of~~``intensity frontier'' experiments initiated in the 1970s. This earlier period, however, was driven partly by studies of hadronic and neutrino physics, and partly by searches for the Higgs boson and the QCD axion. Indeed, in its original incarnation, the QCD axion was part of the electroweak Higgs sector and had its mass spanning the range of $\OO$(100~keV--1~MeV) \cite{Peccei:1977hh,Peccei:1977ur,Weinberg:1977ma,Wilczek:1977pj}. With increasing constraints and no discoveries, laboratory searches for the QCD axion withered away in the early 1990s. By then the consensus was that the Peccei-Quinn (PQ) mechanism had to take place at much higher energy scales, resulting in the ``invisible'' axion \cite{Kim:1979if,Shifman:1979if,Dine:1981rt}. The tradeoff for foregoing PQ symmetry breaking at the electroweak scale was an ultralight axion with the correct cosmological relic abundance to explain dark matter \cite{Preskill:1982cy,Abbott:1982af,Dine:1982ah}, which has been the focus of several ongoing and proposed experiments \cite{Aalseth:2002qf,Asztalos:2009yp,Budker:2013hfa,Lamoreaux:2013koa,Armengaud:2014gea,Arvanitaki:2014dfa,Rybka:2014cya,Graham:2015ouw,Kahn:2016aff, Baryakhtar:2018doz,Irastorza:2018dyq,Brun:2019lyf}.

Nonetheless, motivations for the scale of PQ symmetry breaking are a matter of theoretical prejudice. 
In the original PQWW\footnote{PQWW stands for Peccei-Quinn-Weinberg-Wilczek.} axion model, a single mechanism to break PQ and electroweak symmetries tackled two major puzzles at once---the absence of CP violation in the strong interactions, 
and the generation of masses for SM particles\footnote{Amusingly, the original axion was allegedly nicknamed \emph{higglet} by Roberto Peccei and Helen Quinn. \emph{Higglet} was also the terminology used by Bill Bardeen and Henry Tye in \cite{Bardeen:1977bd}.}. Conversely, axion models with high PQ breaking scales $f_\text{PQ}\gsim 10^9$~GeV could simultaneously address the strong CP problem and the origin of dark matter. In this \emph{article}, we focus on yet another possibility, whereby the PQ mechanism is realized by new dynamics close to the QCD scale. Considering that the solution to the strong CP problem provided by the PQ mechanism is intimately connected with the nonperturbative dynamics of QCD, it is not farfetched to suppose that their scales should not be separated by over 10 orders of magnitude. Indeed, such wide separation of scales makes the delicate cancellation mechanism of the strong CP phase vulnerable to spoiling effects, such as nonperturbative quantum gravity effects, which, based on general arguments, are expected to violate global symmetries \cite{Holman:1992us,Kamionkowski:1992mf,Barr:1992qq,Ghigna:1992iv, Kallosh:1995hi} (see also \cite{Gori:2020xvq} for a shared point of view). Furthermore, existing anomalies in nuclear transitions \cite{Krasznahorkay:2015iga,Krasznahorkay:2019lyl} and in the $\pi^0$ decay width to $e^+e^-$ \cite{Abouzaid:2006kk}, if confirmed as beyond the Standard Model (BSM) phenomena, would strongly support the possibility of a low PQ scale axion, as we shall discuss.

A light BSM sector realizing the PQ mechanism at a scale of $\OO$(GeV) cannot be completely generic, however. Any new degrees of freedom must either have weak or non-generic couplings to avoid existing experimental constraints (which is the case of the \emph{electrophilic}, \emph{muophobic} and \emph{piophobic} QCD axion studied in \cite{Alves:2017avw}), or they must have predominantly hadronic couplings and ``blend in'' with the QCD resonances in the spectral range of $\sim400~\text{MeV} - 2~\text{GeV}$. The phenomenology of the latter is quite challenging to predict and to probe experimentally. On the other hand, the inevitable pseudo-Goldstone degree of freedom, manifested as the QCD axion, is much more amenable to phenomenological studies using Chiral Perturbation Theory ($\chiPT$). Indeed, a robust prediction of $\chiPT$ is that the mass of the QCD axion should lie in the range $m_a\sim$ 1 -- 20~MeV when its decay constant is $f_a\sim\OO$(1 -- 10)~GeV. For generic models in this range, the axion mixing angle with the neutral pion is quite large, $\thetaPI\sim\OO(f_\pi/f_a)\sim\OO(0.01-0.1)$, and strongly excluded by bounds on rare pion decays, which require $\thetaPI\lsim\OO(10^{-4})$. However, as shown in \cite{Alves:2017avw}, axion-pion mixing can be suppressed well below its generic magnitude if the axion couples exclusively with light quarks, $u$ and $d$, with PQ-charge assignments $\qu=2\,\qd$. In this special region of parameter space, the phenomenology of the QCD axion is no longer dominated by its isovector couplings; instead, it is largely determined by its isoscalar mixings with the $\eta$ and $\eta^\prime$ mesons. As such, it inherits the same strong dependence of the $\eta$ and $\eta^\prime$ on higher order terms in the chiral expansion, 
and its hadronic couplings suffer from $\OO(1)$ uncertainties.

Despite these large uncertainties stemming from $\chiPT$, it is still possible to parametrize the dependence of a variety of hadronic signatures of the axion in terms of its isovector and isoscalar mixing angles, while remaining agnostic about their magnitudes. The usefulness of such parametrization is manifest when confronting experimental data, not only in constraining the axion's hadronic mixing angles, but also in interpreting experimental anomalies as potential signals of the QCD axion. This will be the underlying philosophy of this study\footnote{This same philosophy was adopted by the authors of \cite{Aloni:2018vki} in the study of hadronically coupled ALPs.}.

Complementary, the underlying motivation for this study is a combination of the longstanding puzzle posed by the strong CP problem, and three independent experimental anomalies. The first two refer to bump-like excesses observed in specific magnetic transitions of $\Be$ and $\He$ nuclei via $e^+e^-$ emission, with (na\"{\i}ve) significances of $6.8\,\sigma$ \cite{Krasznahorkay:2015iga} and $7.2\,\sigma$ \cite{Krasznahorkay:2019lyl}, respectively. The third anomaly is related to the persistently high central value observed for the width $\Gamma(\pi^0\to e^+e^-)$, whose most recent and precise measurement, performed by the KTeV Collaboration in 2007 \cite{Abouzaid:2006kk}, showed a discrepancy from the theoretical expectation in the SM at the level of $\sim2-3.2\,\sigma$ \cite{Dorokhov:2007bd,Dorokhov:2008qn,Vasko:2011pi,Husek:2014tna}. In combination, these anomalies point to a common BSM origin: a new short-lived boson with mass of $\sim16-17$ MeV, coupled to light quarks and electrons, and decaying predominantly to $e^+e^-$ (see also \cite{Kirpichnikov:2020tcf} for connections with other anomalies). As an \emph{ad hoc} explanation, there are only two possibilities for the spin and parity of this hypothetical new boson: it can either be a pseudoscalar ($J^P=0^-$), or an axial-vector ($J^P=1^+$), in order to simultaneously account for these three excesses\footnote{In particular, the $1^-$ \emph{protophobic} vector boson proposed by Feng \emph{et.\,al} in \cite{Feng:2016jff,Feng:2016ysn} as an explanation of the $\Be$ anomaly cannot be emitted in the $0^-\to 0^+$ transition of $\He$, nor does it contribute non-negligibly to $\Gamma(\pi^0\to e^+e^-)$. In \cite{Feng:2020mbt}, Feng \emph{et.\,al} proposed an alternative explanation of the $\He$ anomaly, whereby the $e^+ e^-$ excess stems from the de-excitation of the overlapping $0^+$ nuclear state. Recently, \cite{Zhang:2020ukq} argued that the \emph{protophobic} vector boson hypothesis is excluded as an explanation of the $\Be$ anomaly.}. Further constraints push these two possibilities into peculiar regions of parameter space, which may require contrived and/or baroque UV completions\footnote{For instance, in the axial-vector case, the model building required to circumvent stringent bounds from electron-neutrino scattering restricts the axial-vector couplings of the $1^+$ state to light quarks to satisfy $g_u^A=-2\,g^A_d$ \cite{Kahn:2016vjr,Kozaczuk:2016nma}; axial-vector models also typically require many \emph{ad hoc} degrees-of-freedom to cancel gauge anomalies in the UV. In the axion case, 
in order to suppress $a-\pi^0$ mixing, the PQ charges of the up and down quarks must satisfy $\qu=2\,\qd$, with (nearly) vanishing PQ charges for the other quarks. 
Such flavor alignment, combined with the fact that $f_{PQ}\sim\OO$(GeV), requires nontrivial UV completion at the weak scale, see \cite{Alves:2017avw}.}. At face value neither of them is particularly compelling, leading many to believe that these anomalies are either the result of experimental systematics and/or poorly understood SM effects. In our opinion, this illustrates the paradoxical predicament of the light dark sector intensity frontier program: the generic models it seeks to discover or rule out are not strongly motivated, and, at least historically, it has been the case that experimental excesses without theoretically compelling interpretations tend to be received with strong skepticism.

Fortunately, this predicament might not be warranted here. Nuclear transitions via axion emission and (modified) rare meson decays are smoking gun signatures of the QCD axion which have been predicted over three decades ago \cite{Treiman:1978ge,Donnelly:1978ty,Barroso:1981bp,Antoniadis:1981zw,Bardeen:1986yb,Krauss:1986bq,Davier:1986ps,Krauss:1987ud}. The fact that some of these signatures have appeared in $\Be$, $\He$, and $\pi^0$ decays, and can be consistently explained by a QCD axion variant which remains experimentally viable (albeit with peculiar properties of \emph{electrophilia}, \emph{muophobia} and \emph{piophobia}), should be taken with 
cautious optimism.

After a brief overview of the most relevant properties of the \emph{piophobic} QCD axion in Sec.\,\ref{overview}, we obtain the parameter space of axion isoscalar couplings favored by the $\Be$ and $\He$ anomalies, and, taking into account nuclear and hadronic uncertainties, show that they significantly overlap, favoring the QCD axion emission hypothesis as a single explanation of both anomalies (Sec.\,\ref{BeHe}). We then turn to axion signals in rare meson decays. In Subsec.\,\ref{secEtaee}, we obtain the the parametric dependence of $\eta/\eta^\prime$ di-electronic decays on the axion's isoscalar mixing angles. In Subsec.\,\ref{secEtaHad}, we calculate the rate for axio-hadronic decays of the $\eta$ and $\eta^\prime$ mesons in the framework of \emph{Resonance Chiral Theory}, an effective ``UV completion'' of $\chiPT$ that incorporates low-lying QCD resonances and extends the principle of \emph{Vector Meson Dominance}. Finally, in Sec.\,\ref{secKaon}, we investigate various axionic decays of charged and neutral Kaons, considering distinct possible implementations of octet enhancement in $\chiPT$ and their effect on axionic Kaon decay rates. We conclude in Sec.\,\ref{conclusion}.

\section{Brief overview of the piophobic QCD axion}
\label{overview}
Generic models of the QCD axion with mass of $\sim 16 - 17$ MeV are largely excluded. However, as investigated in \cite{Alves:2017avw}, all experimental constraints to date can be avoided in this mass range if the axion satisfies a few specific requirements:
\begin{itemize}
\item[(i)]{it must be short-lived$^{\ref{oi}}$ ($\tau_a\lesssim 10^{-13}$ s), and decay predominantly to $e^+e^-$ in order to avoid limits from beam dump and fixed target experiments, as well as constraints from charged Kaon decays such as $K^+\to \pi^+ (a\to \gamma\gamma\,,\;\text{invisible})$;}
\item[(ii)]{the PQ charges of $2^{nd}$ and $3^{rd}$ generation SM fermions must vanish or be suppressed, in order to avoid limits from the muon anomalous magnetic dipole moment, $(g-2)_{\mu}$, and from upper bounds on radiative quarkonium decays: $J/\Psi,\Upsilon\to \gamma\, (a\to e^+e^-)$;}
\item[(iii)]{the $a-\pi^0$ mixing must be suppressed, $\thetaPI\lsim\OO(10^{-4})$, in order to respect upper bounds on $\Br\big(\pi^+\to e^+\nu_e (a\to e^+e^-)\big)$.}
\end{itemize}

A simple phenomenological IR model realizing the requirements above can be easily incorporated in the post-electroweak symmetry breaking SM Lagrangian by ascribing axionic phases to the masses of the up-quark, down-quark, and electron:
\beqnarray\label{axionicmasses}
m_u\,&\to&\,m_u\,e^{i\,\gamma^5\,\qu\;a/f_a}\,,\nonumber\\
m_d\,&\to&\,m_d\,e^{i\,\gamma^5\,\qd\;a/f_a}\,,\\
m_e\,&\to&\,m_e\,e^{i\,\gamma^5\,\qe\;a/f_a}\,,\nonumber
\eeqnarray
where $q^f_{\text{\tiny PQ}}$ ($f=u,d,e$) are PQ charges, with $\qe\sim\OO(1)$ and $\qu=2\,\qd$\,. Importantly, no additional operators should be present in this specific basis, such as derivative couplings of the axion to quark axial-currents, or the usual linear coupling of the axion to the gluon dual field strength operator.

In this IR model, requirement (ii) mentioned above has been imposed by \emph{fiat}. Requirement (i) follows from the axion's coupling to $e^+e^-$, which dominates its decay width:
\beqnarray
\Gamma(a\to e^+e^-)~&=&~\frac{m_a}{8\pi}\,\left(\frac{\qe\,m_e}{f_a}\right)^2\,\sqrt{1-\frac{4\,m_e^2}{m_a^2}\,}\,,\label{awidth}\\
\Rightarrow\qquad \tau_a~&\approx&~\frac{\;4\times 10^{-15}~\text{s}\;}{(\qe)^2}\,.\label{alifetime}
\eeqnarray
For $m_a\sim16-17$ MeV, existing bounds on the electron's PQ charge are very mild, limiting its range to $1/5\lsim|\qe|\lsim2$. The upper bound is set by KLOE's 2015 search for visibly decaying dark photons \cite{Anastasi:2015qla}, whereas the lower bound is set by the 2019 results from CERN's SPS NA64 fixed target experiment\footnote{\label{oi} See added note at the end of this article.} \cite{Banerjee:2019hmi,Depero:2020zfy}, constraining the axion lifetime to $\tau_a\lsim 10^{-13}$\;s. The sensitivities of future experiments to the axion's electronic couplings (such as fixed targets and $e^+e^-$ colliders) have been explored in \cite{Alves:2017avw}.

From (\ref{axionicmasses}) and standard $\chiPT$ at leading order, the axion mass is given by:
\beq
m_a~=~\frac{\left|\,\qu+\qd\,\right|}{\sqrt{\,1+\epsilon_s\,}}\frac{\sqrt{m_u\,m_d}}{(m_u+m_d)}\frac{m_\pi\,f_\pi}{f_a}\,,
\eeq
with
\beq
\epsilon_s~\approx~\frac{m_u\,m_d}{(m_u+m_d)^2}\,\frac{m_\pi^2}{m_K^2}\left(1+6\,\frac{m_K^2}{m^2_{\eta^\prime}}\right)~\simeq~0.04\,.
\eeq

It follows then that for $\qu/2=\qd=1$ and $m_a=16.7$ MeV, the axion decay constant is $f_a\simeq1030$ MeV. We will benchmark $m_a$, $f_a$, $\qu$, and $\qd$ to these values for the remainder of this article.

For generic parameter space of QCD axion models, the quark mass hierarchy $m_{u,d}\ll m_s$ typically induces a hierarchy of axion-meson mixing angles, $\thetaPI\gg\theta_{a\eta}\,,\theta_{a\eta^\prime}$, resulting in the isovector couplings of the axion dominating its experimental signatures. This is not the case for the \emph{piophobic} axion we are considering. Here, the $a-\pi^0$ mixing angle, to leading order in $\chiPT$, is given by:
\begin{widetext}
\beq\label{apimix}
\thetaPI|_{_{\chiPT\,\text{LO}}}~=~-~\frac{f_\pi}{f_a}\bigg(\frac{(m_u\,\qu-m_d\,\qd)}{m_u+m_d}~+~\epsilon_s\,\frac{(\qu-\qd)}{2}\bigg)\,\frac{1}{1+\epsilon_s}
\eeq
\end{widetext}
which, after taking $\qu/2=\qd=1$ and $m_u/m_d=0.485\pm0.027$ from \cite{Fodor:2016bgu}, results in:
\beq\label{Napimix}
\thetaPI|_{_{\chiPT\,\text{LO}}}~=~(-0.02\pm3)\times 10^{-3}\,.
\eeq
It is clear from (\ref{apimix}) and (\ref{Napimix}) that the axion's \emph{piophobia} is the result of an accidental cancellation in $\chiPT$'s leading order contribution to $\thetaPI$. This cancelation stems from the near numerical coincidence  between $m_u/m_d$ and $\qd/\qu = 1/2$.

Unfortunately, 
$\chiPT$'s prediction (\ref{Napimix}) alone is not precise enough to be useful. We instead have resort to observation to determine the allowed range for $\thetaPI$ with better precision. This can be achieved by requiring that the $3.2\,\sigma$ excess in KTeV's measurement of $\Gamma(\pi^0\to e^+e^-)$ \cite{Abouzaid:2006kk} be the result of $\pi^0-a$ mixing, which yields \cite{Alves:2017avw}:
\beq\label{apimixKTeV}
\thetaPI|_{_{\text{KTeV}}}~=~\frac{(-0.6\pm0.2)}{\qe}\times 10^{-4}\,.
\eeq 

Given the suppressed value (\ref{apimixKTeV}) for $\thetaPI$, this model features an atypical hierarchy of mixing angles, $\thetaPI\ll\theta_{a\eta}\,,\theta_{a\eta^\prime}$, which results in the isoscalar couplings of the axion dominating its experimental signatures. This aggravates the loss of $\chiPT$'s usual predictive power in axion phenomenology---given its state-of-the-art, $\chiPT$ cannot numerically pin down the isoscalar mixing angles $\theta_{a\eta}\,,\theta_{a\eta^\prime}$ with good accuracy. As argued in \cite{Alves:2017avw}, $\theta_{a\eta}\,,\theta_{a\eta^\prime}$ receive $\OO(1)$ contributions from operators at $\OO(p^4)$ in the chiral expansion, many of which have poorly determined Wilson coefficients.

Any substantive theoretical progress in better determining the axion's hadronic couplings is unlikely to be accomplished anytime soon. Indeed, such efforts might be superseded by future experimental results which will be able to either exclude or narrow down the preferred ranges for the axion's isoscalar couplings. With this in mind, in this study we choose to remain agnostic about their magnitude, and instead simply parametrize the \emph{physical} axion current as \footnote{We omit the dependence of (\ref{Japhys}) on $\bar{e}\gamma_\mu\gamma_5e$, which has no bearing on the axion-meson mixing angles.}:
\begin{widetext}
\beq\label{Japhys}
J_\mu^{\,a_\text{phys}}\;\equiv\;  f_a\, \partial_\mu a_\text{phys}\; \equiv\; \frac{f_a}{f_\pi}\left(f_\pi\,\partial_\mu a \,+\, \thetaPI\, J_{5\,\mu}^{\,(3)} \,+\, \theta_{a\eta_{ud}}\, J_{5\,\mu}^{\,(ud)} \,+\, \theta_{a\eta_{s}}\,J_{5\,\mu}^{\,(s)}\,\right) \,,
\eeq
\end{widetext}
where
\begin{subequations}\label{Jchi}
\begin{alignat}{3}
J_{5\,\mu}^{\,(3)} &~\equiv~ \frac{\bar{u}\gamma_\mu\gamma_5u - \bar{d}\gamma_\mu\gamma_5d}{2} &&~\equiv~f_\pi\, \partial_\mu\pi_3 \,,
\label{J3}\\
J_{5\,\mu}^{\,(ud)} &~\equiv~ \frac{\bar{u}\gamma_\mu\gamma_5u + \bar{d}\gamma_\mu\gamma_5d}{2} &&~\equiv~ f_\pi\, \partial_\mu\eta_{ud} \,,
\label{J8}\\
J_{5\,\mu}^{\,(s)} &~\equiv~ \frac{~\bar{s}\gamma_\mu\gamma_5s~}{\sqrt{2}} &&~\equiv~\, f_\pi\,  \partial_\mu\eta_s \,.
\label{J0}
\end{alignat}
\end{subequations}
The axionic field $a$ and the neutral meson degrees of freedom $\pi_3$, $\eta_{ud}$, and $\eta_s$ in (\ref{Jchi}) mix amongst themselves to yield the physical degrees of freedom (\ie, the mass eigenstates) $a_\text{phys}$, $\pi^0$, $\eta$, and $\eta^\prime$. In particular, the implication of (\ref{Japhys}) is that any strong or weak process involving the currents in (\ref{Jchi}) will have a corresponding axion signature for which one of the neutral mesons in the amplitude gets replaced by $a_\text{phys}$ properly weighted by the appropriate mixing angle.

With the parametrization in (\ref{Japhys}), it is straightforward to obtain the axion's couplings to photons and nucleons. Specifically, below the QCD confinement scale, the electromagnetic anomaly of the physical axion current (\ref{Japhys}) leads to:
\beq\label{aFFdual}
\mathcal{L}_a~\supset~\frac{\alpha}{4\pi f_\pi}\,\left(\thetaPI+\frac{5}{3}\,\thetaUD+\frac{\sqrt{2}}{~3}\,\thetaS\right)\,a\; F_{\mu\nu}\tilde{F}^{\mu\nu}\,,
\eeq
which, combined with (\ref{awidth}), yields the axion decay width and branching ratio to two photons:
\beqnarray
&&\Gamma(a\to\gamma\gamma)~=~\nonumber\\
&&~~~~\left(\thetaPI+\frac{5}{3}\,\thetaUD+\frac{\sqrt{2}}{~3}\,\thetaS\right)^2\left(\frac{\alpha}{4\pi f_\pi}\right)^2\,\frac{m_a^3}{4\pi}\,,~~~~~\\
\nonumber\\
\Rightarrow~&&\Br(a\to\gamma\gamma)~\approx~\nonumber\\
&&~~~~10^{-7}\times\frac{1}{(\qe)^2}\left(\frac{\thetaPI+\frac{5}{3}\,\thetaUD+\frac{\sqrt{2}}{~3}\,\thetaS}{10^{-3}}\right)^2\,.\label{BRaTogg}
\eeqnarray

The axion's contribution to $(g-2)_e$ stemming from its couplings to electrons and photons has been worked out in \cite{Alves:2017avw}.

Finally, expressing the axion nuclear couplings generically as:
\beq
\mathcal{L}_{aNN}~=~ a~\overline{N}\,i \gamma^5 \Big(g_{aNN}^{(0)} \,+\,g_{aNN}^{(1)}\tau^3\Big) N\,,
\eeq
the parametrization in (\ref{Japhys}) yields the following isovector and isoscalar axion-nucleon couplings, respectively:
\begin{subequations}\label{gaNNDeltaq}
\begin{alignat}{1}
g_{aNN}^{(1)}&~=~\thetaPI\,g_{\pi NN}~=~\thetaPI\,(\Delta u - \Delta d)\,\frac{m_N}{f_\pi}\,,\label{gaN1}\\
g_{aNN}^{(0)}&~=\,\Big(\thetaUD(\Delta u + \Delta d) \;+\; \sqrt{2}\,\thetaS\Delta s\,\Big)\,\frac{m_N}{f_\pi}\,.~~~~\label{gaN0}
\end{alignat}
\end{subequations}
Above, $N$ is the nucleon isospin doublet, $m_N$ is the nucleon mass, and $\Delta q$ quantifies the matrix elements of quark axial-currents in the nucleon via $2\,s_\mu\,\Delta q = \langle N| \bar q\gamma_\mu\gamma_5 q\,|N \rangle$, with $s_\mu$ the nucleon spin-vector. The combination in (\ref{gaN1}) is well determined from neutron $\beta$ decay,
\beq\label{gA}
\Delta u - \Delta d\;=\,g_{A}\,\simeq\;1.27\,.
\eeq
On the other hand, estimations for $\Delta u + \Delta d$ and $\Delta s$ based on data from semi-leptonic hyperon decays, proton deep inelastic scattering, and lattice calculations vary widely \cite{Jaffe:1989jz, Savage:1996zd, Karliner:1999fn, Mallot:1999qb, Leader:2000dw, Cheng:2012qr, QCDSF:2011aa, Engelhardt:2012gd, Abdel-Rehim:2013wlz, Bhattacharya:2015gma, Abdel-Rehim:2015owa, Abdel-Rehim:2015lha, Green:2017keo}, ranging from:
\beq\label{DuDdDs}
0.09\lesssim~\Delta u + \Delta d~\lesssim0.62~~~ \text{and}~~-0.35\lesssim \Delta s\lesssim 0.
\eeq

In the following, we will use (\ref{gaNNDeltaq}) to fit the recent $\Be$ and $\He$ anomalies, and (\ref{Japhys}) to obtain various rare meson decays.

\section{Nuclear Transitions}
\label{BeHe}

One of the smoking gun signatures of axions in the mass range $\OO(\text{keV}-\text{MeV})$ are magnetic nuclear de-excitations via axion emission \cite{Treiman:1978ge,Donnelly:1978ty,Barroso:1981bp}. Indeed, such signals have been extensively searched for during the 1980s \cite{Calaprice:1979pe, Savage:1986ty, Hallin:1986gh, DeBoer:1986cm, Hallin:1987uc,Avignone:1988bv,Savage:1988rg, Datar:1988ju, Freedman:1989gz, Asanuma:1990rm}. However, since the energy of typical nuclear transitions ranges from a few keV to a few MeV, past searches did not place meaningful bounds on axions heavier than $m_a\,\gsim\,2$ MeV.

Recently, the MTA Atomki Collaboration led by A. Krasznahorkay reported on the observation of bump-like excesses in the invariant mass distribution of $e^+e^-$ pairs emitted in the de-excitation of specific states of $\Be$ and $\He$ nuclei \cite{Krasznahorkay:2015iga,Krasznahorkay:2019lyl}. The energy difference $\Delta E$ between the nuclear levels involved in these particular transitions is atypically high, {\it a priori} allowing on-shell emission of particles as heavy as $\sim17-18$~MeV. Furthermore, consistent with the allowed values of angular momentum and parity carried away by the axion 
($J^P = 0^-,\,1^+,\,2^-,\,3^+,\,...$), these excesses appeared in \emph{magnetic} (but not \emph{electric}) transitions. Also, consistent with the emission of a \emph{piophobic} axion, these excesses were observed only in predominantly \emph{isoscalar} (but not \emph{isovector}) transitions.  Axion emission rates for the magnetic dipole transitions of $\Be$ have already been worked out in \cite{Alves:2017avw}; we briefly review the main results here to make this section self-contained. We then estimate the expected axion emission rate for the magnetic monopole transition of $\He$ investigated by Krasznahorkay {\it et al.}, and show that the reported excess rates for both nuclei favor the same range of axion isoscalar mixing angles.

\hspace{1cm}
\subsection{Evidence for the QCD axion in $\Be$ transitions}

In \cite{Krasznahorkay:2015iga}, the MTA Atomki experiment selectively populated specific excited states of the $\Be$ nucleus by impinging a beam of protons with finely-tuned energy on a $^7\text{Li}$ target. They then measured the energy and angular correlation of $e^+e^-$ pairs emitted in de-excitations of these states to the ground state of $\Be$. From these measurements they were able to reconstruct final state kinematic variables, such as the invariant mass of the $e^+e^-$ pair, $m_{e^+e^-}$.
The nuclear levels of interest de-excited to the ground state via magnetic dipole (M1) transitions:
\begin{subequations}\label{eeBe}
\begin{alignat}{1}
&^{\,8}\text{Be}^{*}(17.64)~\to~\,^{\,8}\text{Be}(0)+e^+e^- ~,\label{eeBe17}\\
&~~\Delta E=17.64~\text{MeV},~~\Delta I\approx1\,,\nonumber\\
&\nonumber\\
&^{\,8}\text{Be}^{*}(18.15)~\to~\,^{\,8}\text{Be}(0)+e^+e^- ~,\label{eeBe18}\\
&~~\Delta E=18.15~\text{MeV}~,~~\Delta I\approx0\,.\nonumber
\end{alignat}
\end{subequations}
Above, $\Be(0)$ is the $J^P=0^+$ isospin-singlet ground state of the $\Be$ nucleus, and $^{\,8}\text{Be}^{*}(17.64)$ and $^{\,8}\text{Be}^{*}(18.15)$ are $J^P=1^+$ excited states, whose isospin quantum numbers are predominantly $I=1$ and $I=0$, respectively, but are nonetheless isospin-mixed:
\begin{subequations}\label{BeTangles}
\begin{alignat}{2}
\!\!\!\! |^{\,8}\text{Be}^{*}(17.64)\,\rangle&~=~&\sin\theta_{1^+}\,|\,I=0\,\rangle \,+\, \cos\theta_{1^+}\,|\,I=1\,\rangle,\label{BeTangles17}\\
\!\!\!\! |^{\,8}\text{Be}^{*}(18.15)\,\rangle&~=~&\cos\theta_{1^+}\,|\,I=0\,\rangle \,-\, \sin\theta_{1^+}\,|\,I=1\,\rangle.\label{BeTangles18}
\end{alignat}
\end{subequations}
Their level of isospin mixing, quantified by $\theta_{1^+}$, was estimated by \emph{ab initio} quantum Monte Carlo techniques \cite{Pastore:2014oda, Feng:2016ysn}, and by $\chi$EFT many-body methods \cite{Kozaczuk:2016nma} to fall in the approximate range $0.18\lesssim \sin\theta_{1^+}\lesssim 0.43$. Following \cite{Feng:2016ysn,Kozaczuk:2016nma} we will consider a narrower range for $\sin\theta_{1^+}$ which more accurately describes the width of the electromagnetic transition $^{\,8}\text{Be}^{*}(18.15) \to\,\Be(0)+\gamma$:
 \beq\label{BeTheta}
 0.30~\leq\; \sin\theta_{1^+}\leq~ 0.35\,.
\eeq

In the MTA Atomki experiment \cite{Krasznahorkay:2015iga}, a bump-like feature in the $m_{e^+e^-}$ distribution of the $\Delta I\approx0$ transition (\ref{eeBe18}) was observed on top of the monotonically falling spectrum expected from SM internal pair conversion (IPC) \cite{Krasznahorkay:2015iga}. A statistical significance of $6.8\,\sigma$ was reported for this deviation relative to the IPC expectation. Additionally, it was claimed in \cite{Krasznahorkay:2015iga} that the excess events were consistent with the emission of an on-shell resonance, generically labeled ``$X$'', with mass of $m_X=(16.7\pm0.35_{\text{stat}}\pm0.5_{\text{syst}})$~MeV, promptly decaying to $e^+e^-$. This excess was later corroborated by the same Collaboration with a modified experimental setup \cite{Krasznahorkay:2019lgi}, with a combined fit yielding a relative branching ratio of:
\beq\label{AtomkiBe18}
\frac{\Gamma_X}{\Gamma_\gamma}\;\Bigg|_{^{\,8}\text{Be}^{*}(18.15)}\approx~~(6\pm1)\times10^{-6}
\eeq
with respect to the radiative $\gamma$ width of this transition, $\Be^{*}(18.15)\to\,\Be(0)+\gamma$, of $\Gamma_\gamma(18.15)\approx (1.9\pm0.4)$~eV \cite{Tilley:2004zz}.

As for the $m_{e^+e^-}$ spectrum of the $\Delta I\approx1$ transition (\ref{eeBe17}), no statistically significant deviation from the IPC expectation was observed. 
References \cite{Feng:2016jff,Kozaczuk:2016nma} inferred a na\"{\i}ve upper bound of
\vspace{0.2cm}
\beq\label{AtomkiBe17}
\frac{\Gamma_X}{\Gamma_\gamma}\;\Bigg|_{^{\,8}\text{Be}^{*}(17.64)}\lsim~~\OO(10^{-6})
\eeq
for the de-excitation rate of $^{\,8}\text{Be}^{*}(17.64)$ via on-shell emission of this hypothetical ``$X(17)$'' resonance.

If it is confirmed that the observed excess originates from new, beyond the SM phenomena, as opposed to nuclear physics effects or experimental systematics, it could indeed be explained by the \emph{piophobic} QCD axion. The prediction for axion emission rates from magnetic dipole nuclear transitions was first worked out by Treiman \& Wilczek \cite{Treiman:1978ge} and independently by Donnelly {\it et al.} \cite{Donnelly:1978ty} back in the late 1970s. For the two transitions in (\ref{eeBe}), the axion-to-photon emission rate is (see also \cite{Barroso:1981bp,Bardeen:1986yb,Savage:1988rg}):
\begin{widetext}
\beq\label{nuclearRateM1}
\frac{\Gamma_a}{\Gamma_\gamma}\;\Bigg|_{\Be^*}\!
=~~\frac{1}{2\pi\alpha}\,\left|\frac{\sum_{I=0,1}~\,g_{aNN}^{(I)}~\big\langle \,I\, \big|\,{\Be^{*}}\big\rangle}{~\sum_{I=0,1}~(\mu^{(I)}-\eta^{(I)})~\big\langle \,I\, \big|\,{\Be^{*}}\big\rangle~}\right|^2\left(1-\frac{m_a^2}{{\Delta E}^{\,2}}\right)^{3/2}\,,
\eeq
\end{widetext}
where $|{\Be^{*}}\rangle$ denotes one of the states in (\ref{eeBe}), and its overlap with the isospin eigenstates $|I\!=\!0\rangle$ and $|I\!=\!1\rangle$ follows from (\ref{BeTangles}). The quantities $\mu^{(0)}=\mu_p+\mu_n=0.88$ and $\mu^{(1)}=\mu_p-\mu_n=4.71$ are, respectively, the isoscalar and isovector nuclear magnetic moments, and $\eta^{(0)}$, $\eta^{(1)}$ parametrize ratios of nuclear matrix elements of convection and magnetization currents  \cite{Avignone:1988bv}. In particular, $\eta^{(0)}=1/2$ due to total angular momentum conservation. The nuclear structure dependent parameter $\eta^{(1)}$, to the best of our knowledge, has not been calculated for $\Be$\,; we therefore conservatively vary $\eta^{(1)}$ in the range:
\beq\label{eta1Be}
-1~\leq~~\eta^{(1)}\big|_{\Be}~~\leq~ 1\,.
\eeq

Combining (\ref{nuclearRateM1}) with (\ref{gaNNDeltaq}), (\ref{gA}), (\ref{eeBe18}), and (\ref{BeTangles18}), we can infer the axion isoscalar mixing angles that yield the observed excess rate (\ref{AtomkiBe18}). For concreteness, we vary $\thetaPI$ within the $1\,\sigma$ range favored by the KTeV anomaly fit, (\ref{apimixKTeV}), while also varying $\qe$ in the range $1/2\leq\qe\leq2$, and the nuclear structure parameters $\theta_{1^+}$ and $\eta^{(1)}$ in the ranges (\ref{BeTheta}) and (\ref{eta1Be}), respectively. We obtain:
\beqnarray\label{BeFavoredIsoTheta}
\!\!\!\!\!\!\! -\big( \thetaUD(\Delta u + \Delta d) +&& \sqrt{2}\,\thetaS\Delta s \big)\Big|_{\Be^*(18.15)}\nonumber\\
&&~\approx~(1.1-6.3)\times 10^{-4}\,.
\eeqnarray
%
%
%

\begin{figure}[t]
 \includegraphics[width=1\columnwidth]{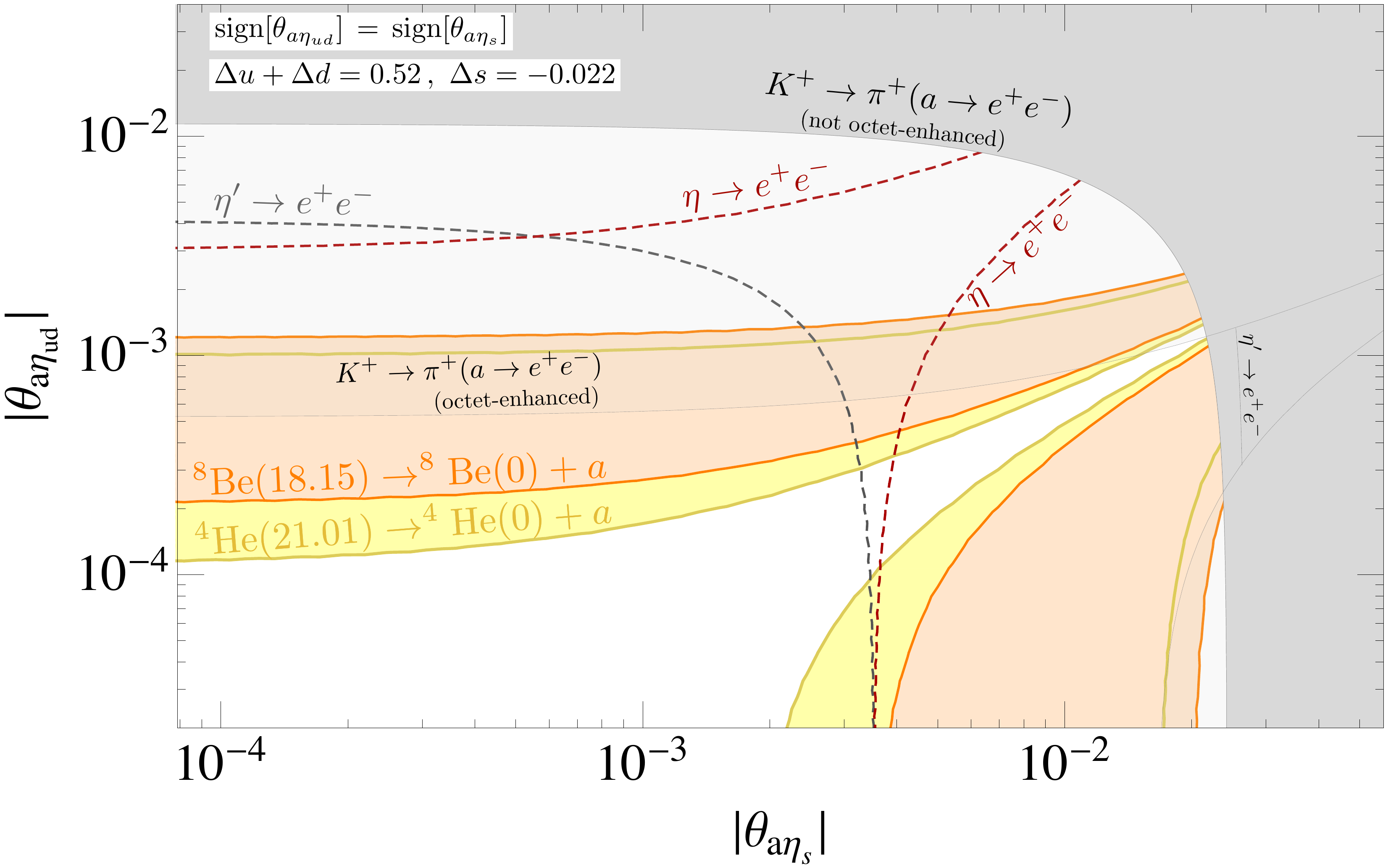}
 \includegraphics[width=1\columnwidth]{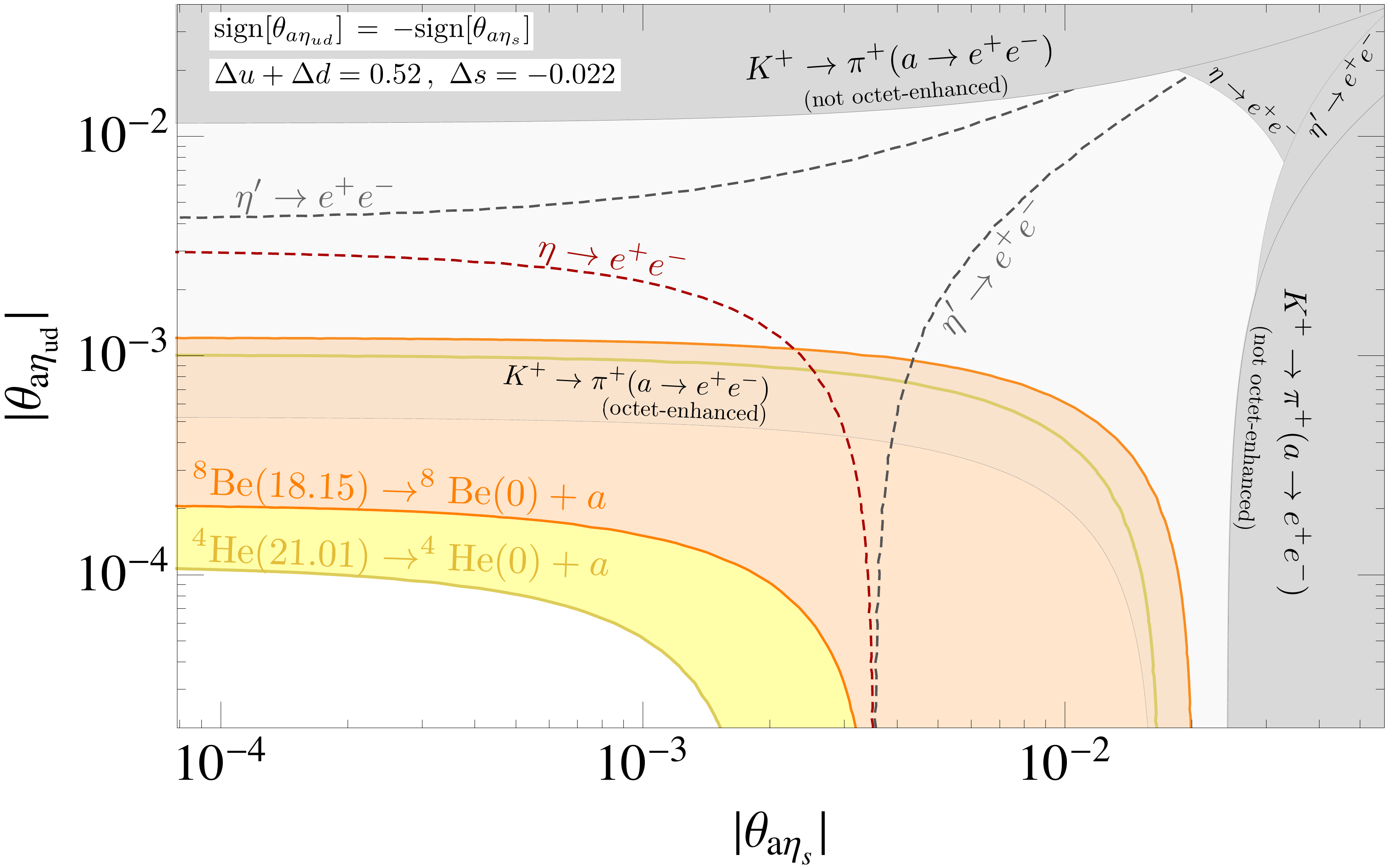}
\caption{\small Fits, constraints, and sensitivity projections in the parameter space of the axion isoscalar couplings. The upper (lower) plot assumes the same (opposite) relative sign between $\thetaUD$ and $\thetaS$. The orange and yellow bands enclose the range of isoscalar mixing angles that can explain the $\Be$ and $\He$ anomalies, respectively, benchmarking $\Delta u+\Delta d$ and $\Delta s$ to the values shown; {\it cf.} (\ref{BeFavoredIsoTheta}, \ref{HeFavoredIsoTheta}). The shaded gray regions are excluded by the conservative upper bound $\Br(K^+\to \pi^+ (a\to e^+e^-))\lsim 10^{-5}$ (under different scenarios for octet enhancement in $\chiPT$) and by current bounds on $\eta^{(\prime)}\to e^+e^-$, assuming $\qe=1/2$; {\it cf.} (\ref{etaTOeeBound}, \ref{etaprimeTOeeBound}). The dashed gray (red) lines show the expected reach from measurements of (or bounds on) $\eta^\prime (\eta)\to e^+e^-$, assuming that future experiments will have sensitivity to the branching ratios predicted in the SM, (\ref{etaTOeeSM}) and (\ref{etaprimeTOeeSM}), with $\OO(1)$ precision.}
\label{8Be4He}
\end{figure}

Figure \ref{8Be4He} displays the parameter space in $\thetaUD$ vs. $\thetaS$ favored by the $\Be$ anomaly (orange bands) under the assumptions of $\Delta u + \Delta d = 0.52$, $\Delta s = -0.022$ \cite{diCortona:2015ldu}, and equal (upper plot) or opposite (lower plot) relative sign between $\thetaUD$ and $\thetaS$. These bands shift non-negligibly as $\Delta u + \Delta d$ and $\Delta s$ are varied within the ranges in (\ref{DuDdDs}).

Finally, we conclude this discussion by using (\ref{nuclearRateM1}) and (\ref{BeFavoredIsoTheta}) to predict the axion emission rate in the $\Delta I\approx 1$ transition (\ref{eeBe17}):
\beq
\frac{\Gamma_a}{\Gamma_\gamma}\;\Bigg|_{\Be^*(17.64)}\approx~~(0.008-1)\times 10^{-6}\,.
\eeq
Indeed, this rate can be down by as much as 2 orders of magnitude below the sensitivity of published results to date, but could potentially be detectable if sufficient statistics is accumulated in this channel.


\subsection{Evidence for the QCD axion in $\He$ transitions}

More recently, the same Collaboration led by A. Krasznahorkay investigated transitions of a different nucleus, $\He$ \cite{Krasznahorkay:2019lyl}. With a 900 keV proton beam bombarding a $^3$H fixed target, this experiment populated the first two excited states of $\He$:
\begin{subequations}\label{4HeStates}
\begin{alignat}{2}
^{\,4}\text{He}^{*}(20.49)~,~~~ J^P = 0^+~,~~~ I=0~,~~~\Gamma = 0.50~\text{MeV}\,,\label{4He20}\\
^{\,4}\text{He}^{*}(21.01)~,~~~J^P = 0^- ~,~~~ I=0~,~~~\Gamma = 0.84~\text{MeV}\,,\label{4He21}
\end{alignat}
\end{subequations}
and similarly measured the emission of $e^+e^-$ pairs from de-excitations of these states to the $I=0$, $J^P=0^+$ ground state, denoted here by $\He(0)$. Such transitions are allowed via:

\begin{subequations}\label{4HeAllowed}
\begin{alignat}{1}
&\!\!\!\!\!\!\!\!\!\text{(E0)}~~~~ ^{\,4}\text{He}^{*}(20.49)\,\to~\He(0)\,+\,(\gamma^*\to e^+e^-)\,,\label{4HeE0g}\\
&~~~~~~~~\Delta E=20.49~\text{MeV}\,,~~\Delta I=0\,,\nonumber\\
&\nonumber\\
&\!\!\!\!\!\!\!\!\!\text{(M0)}~~~~ ^{\,4}\text{He}^{*}(21.01)\,\to~\He(0)\,+\,(a~\to e^+e^-)\,,\label{4HeM0a}\\
&~~~~~~~~\Delta E=21.01~\text{MeV}\,,~~\Delta I=0\,,\nonumber
\end{alignat}
\end{subequations}
but \emph{forbidden} to occur via the following processes:

\begin{subequations}\label{4HeForbidden}
\begin{alignat}{1}
&\!\!\!\!\!\!\!\!\!\text{(E0)}~~~~ ^{\,4}\text{He}^{*}(20.49)\,\not\to~\He(0)\,+\,(a\;\to\,e^+e^-)\,,\label{4HeE0a}\\
&\nonumber\\
&\!\!\!\!\!\!\!\!\!\text{(M0)}~~~~ ^{\,4}\text{He}^{*}(21.01)\,\not\to~\He(0)\,+\,(\gamma^*\to e^+e^-)\,.\label{4HeM0g}
\end{alignat}
\end{subequations}
Above, E0 and M0 refer, respectively, to the electric monopole ($J^{P}=0^+$) and magnetic monopole ($J^{P}=0^-$) multipolarities of these transitions.

After cuts, background subtraction, and accounting for contributions to the $m_{e^+e^-}$ spectrum from (\ref{4HeE0g}) and from external pair conversion originating from the radiative proton capture reaction $^3\text{H}(p,\gamma)\He$, a suggestive bump-like excess was observed in the final $m_{e^+e^-}$ distribution, with a statistical significance of $7.2\,\sigma$. Under the assumption that this excess originated from on-shell emission of a narrow resonance from the M0 transition (\ref{4HeM0a}), the fit to the data performed in \cite{Krasznahorkay:2019lyl} yielded a favored resonance mass of $m_a=(16.84\,\pm\,0.16_{\text{stat}}\,\pm\,0.20_{\text{syst}})$~MeV, and de-excitation width\footnote{No error bars were provided for (\ref{Gamma4Hea}) in \cite{Krasznahorkay:2019lyl}.}:
\beq\label{Gamma4Hea}
\Gamma\big|_{{\He^*}(21.01)\;\to\;{\He}(0)\,+\,a}~~\approx~~3.9\times10^{-5}~\text{eV}\,.
\eeq

It is encouraging that not only the same resonance mass (within error bars) is favored by fits to both the $\He$ and $\Be$ excesses, but also that they appear in \emph{magnetic} and (dominantly) \emph{isoscalar} transitions, compatible with the interpretation of \emph{piophobic} axion emission. To further support this hypothesis, we must obtain the range of axion isoscalar mixing angles compatible with the observed rate. According to Donnelly {\it et al.} \cite{Donnelly:1978ty}, the width of axionic emission in $0^-\!\to\, 0^+$ nuclear transitions is estimated to be\footnote{In \cite{Feng:2020mbt}, the calculated rate for pseudoscalar emission in this $0^-\to 0^+$ transition assumed a nonderivatively-coupled pseudoscalar ``$X$'' (see the effective operator in eq.\,(39) of \cite{Feng:2020mbt}), resulting in an amplitude with no momentum dependence (eq.\,(49) of \cite{Feng:2020mbt}) and an emission rate scaling as $\Gamma_X\propto |\vec{p}_X|$. Under this assumption, the authors of \cite{Feng:2020mbt} concluded that the rate of pseudoscalar emission in this transition would be six orders of magnitude larger than the experimentally favored rate. We point out that their conclusion hinged on their assumption that the leading effective operator at the nuclear level mediating this transition was a relevant operator of dimension 3. In the case of the QCD axion, this assumption is not valid, since the axion only couples derivatively to nuclear axial currents. For the QCD axion, the leading effective nuclear operator is dimension 5, resulting in
an amplitude scaling as $\propto |\vec{p}_a|^2$, and therefore an emission rate scaling as $\Gamma_a\propto |\vec{p}_a|^5$. Note that the axionic amplitude is still isotropic, as it should be for a monopole transition, despite its nontrivial momentum dependence. For details, see \cite{Donnelly:1978ty,Donnelly:1978tz, Barroso:1981bp}.}:
\beq\label{DonM0}
\Gamma_a\Big|_{\text{M0}}\approx~\frac{2}{\,(2\,I_{N^*}+1)\,}\,\frac{|\vec{p}_a|^5}{m_N^2\, Q^2}~\big|\,a_{\text{\tiny M0}}^{(0)}~ g_{aNN}^{(0)}\,+\, a_{\text{\tiny M0}}^{(1)}~ g_{aNN}^{(1)} \, \big|^2\,,
\eeq
where $I_{N^*}$ is the isospin of the excited nuclear state, $|\vec{p}_a|\approx\sqrt{\Delta E^2-m_a^2\,}$ is the magnitude of the axion's spatial-momentum in the rest frame of the decaying nucleus, $Q$ is a typical nuclear momentum transfer (of order the nucleus Fermi momentum, $Q\approx k_F\approx 250$ MeV), and $a_{\text{\tiny M0}}^{(0)}$, $a_{\text{\tiny M0}}^{(1)}$ involve nuclear matrix elements of magnetization currents, and are of $\OO(1)$, unless forbidden by isospin conservation. For an isoscalar transition such as (\ref{4HeM0a}), (\ref{DonM0}) reduces to:
\beq\label{Don4HeM0}
\Gamma_a\Big|_{\text{M0},\;\Delta I=0}~\approx~~\big|a_{\text{\tiny M0}}^{(0)} \big|^2 ~\frac{\,2\,(\Delta E^2-m_a^2)^{5/2}\,}{m_N^2\, Q^2}~\big|g_{aNN}^{(0)}\big|^2\,.
\eeq

Using (\ref{Don4HeM0}), (\ref{gaN0}), and varying $a_{\text{\tiny M0}}^{(0)}$ in the range $1/3\leq |a_{\text{\tiny M0}}^{(0)}| \leq 3$, we find that the axionic de-excitation width of the M0 transition (\ref{4HeM0a}) in $\He$ yields the observed rate (\ref{Gamma4Hea}) if:
\beqnarray\label{HeFavoredIsoTheta}
\!\!\!\!\!\!\!  -\big( \thetaUD(\Delta u + \Delta d) +&& \sqrt{2}\,\thetaS\Delta s \big)\Big|_{\He^*(21.01)}\nonumber\\
&&~\approx~(0.58-5.3)\times 10^{-4}\,,
\eeqnarray
which is compatible with the range of axion isoscalar mixing angles favored by the $\Be$ excess, (\ref{BeFavoredIsoTheta}). In Fig.\,\ref{8Be4He} we likewise display the parameter space in $\thetaUD$ vs. $\thetaS$ favored by the $\He$ anomaly (yellow bands) under the same assumptions for $\Delta u + \Delta d$, $\Delta s$ and relative sign between $\thetaUD$ and $\thetaS$ used in the computation of the $\Be$ orange bands. The $\He$ yellow bands also shift non-negligibly as $\Delta u + \Delta d$ and $\Delta s$ are varied within the ranges in (\ref{DuDdDs}).

It is remarkable that the \emph{piophobic} QCD axion is able to \emph{simultaneously} explain the reported rate of anomalous excesses in de-excitations of two very different nuclei, $\Be$ and $\He$, as shown by the overlap between the favored ranges for the axion isoscalar mixing angles (\ref{BeFavoredIsoTheta}) and (\ref{HeFavoredIsoTheta}), or, equivalently, by the overlap between the yellow and orange bands in Fig.\,\ref{8Be4He}. This weakens the case for a nuclear physics origin of the observed features in the $m_{e^+e^-}$ spectra of these transitions \cite{Zhang:2017zap}. And the fact that ``unexplained'' features are absent in the $m_{e^+e^-}$ spectrum of several other measured transitions---(\ref{eeBe17}) being one example---also makes it less straightforward to ``explain away'' the observed excesses
as poorly understood experimental systematics. We therefore reiterate our point, stated in the \emph{Introduction} (Sec.\,\ref{intro}), that the anomalies in $\Be$ and $\He$ transitions, and their quantitative compatibility with predicted signals from the QCD axion, should not be quickly dismissed. A cautiously optimistic attitude and support for an independent verification of these measurements are certainly warranted, as well as further exploration of other isoscalar magnetic nuclear transitions with $\Delta E \gsim 17\;$MeV, and radiative $\pi^-/p/n$ capture reactions with significant magnetic components \cite{Chen:2019ivz}.

\section{$\eta$ and $\eta^\prime$ decays}
\label{eta}

In light of the anomalies in nuclear de-excitations discussed in the previous section, a natural next step is to investigate other systems where the hadronic couplings of the \emph{piophobic} QCD axion could be more precisely determined or more stringently constrained. In this section we consider rare decays of $\eta$ and $\eta^\prime$ mesons, which, with the prospect of future $\eta/\eta^\prime$ factories, could become powerful future probes of axions and hadronically coupled ALPs more generally \cite{Gan:2020aco}. These include the second phase of the JLab Eta Factory (JEF) program \cite{Mack:2014gma}, expected to improve existing bounds on rare $\eta$ decays by two orders of magnitude, and the REDTOP experiment \cite{Gonzalez:2017fku,Gatto:2016rae}, a planned $\eta/\eta^\prime$ factory projected to deliver as many as $10^{13}~\eta$ mesons and $10^{11}~\eta^\prime$ mesons. These will offer an unprecedented opportunity to study rare $\eta/\eta^\prime$ decays and probe BSM physics.

\vspace{-0.2cm}
\subsection{Di-electronic $\eta$ and $\eta^\prime$ decays}
\label{secEtaee}
Just as the precise KTeV measurement of $\pi^0\to e^+e^-$ offered the best determination of $\thetaPI$, future observations of $\eta\to e^+e^-$ and $\eta^\prime\to e^+e^-$ could narrow down the ranges for the axion isoscalar mixing angles $\thetaUD$ and $\thetaS$. Present bounds on these dileptonic branching ratios \cite{Agakishiev:2013fwl,Achasov:2018idb} are still two orders of magnitude away from sensitivity to the predicted SM rate \cite{Landsberg:1986fd,Dorokhov:2009xs,Masjuan:2015cjl}:
\begin{subequations}\label{etaTOee}
\begin{alignat}{2}
\Br(\eta~\to~e^+e^-)_\text{exp}~~&<&~~7\times 10^{-7}\,,\label{etaTOeeBound}\\
\Br(\eta~\to~e^+e^-)_\text{SM}~~&\approx&~~(4.6-5.4)\times 10^{-9}\,,\label{etaTOeeSM}
\end{alignat}
\end{subequations}
and
 \begin{subequations}\label{etaprimeTOee}
\begin{alignat}{2}
\Br(\eta^\prime~\to~e^+e^-)_\text{exp}~~&<&~~0.56\times 10^{-8}\,,\label{etaprimeTOeeBound}\\
\Br(\eta^\prime~\to~e^+e^-)_\text{SM}~~&\approx&~~(1-2)\times 10^{-10}\,.\label{etaprimeTOeeSM}
\end{alignat}
\end{subequations}
Indeed, the highly suppressed SM contribution to these dileptonic channels makes them potentially sensitive to a variety of interesting new physics scenarios.

In anticipation of a future discovery of these decay modes, we obtain the axionic contribution to the dileptonic decays $\eta^{(\prime)}\to e^+e^-$ due to $a-\eta^{(\prime)}$ mixing. Assuming that this effect dominates these rates ({\it i.e.}, that interference with the SM amplitudes can be neglected), we have: 
\beq
\Gamma\big(\eta^{(\prime)}\to e^+e^-\big)~\approx~\frac{m_{\eta^{(\prime)}}}{8\pi}\,\left(\frac{\qe\,m_e}{f_a}~\theta_{a\eta^{(\prime)}}\!\!\right)^2\sqrt{1-\frac{4\,m_e^2}{~m_{\eta^{(\prime)}}^2}}\,.
\eeq

The mixing angles $\theta_{a\eta}$ and $\theta_{a\eta^{\prime}}$ can be re-expressed in terms of $\thetaUD$ and $\thetaS$ using the parametrization \cite{Escribano:2005qq}
\begin{subequations}\label{etaparamet}
\begin{alignat}{2}
|\eta\,\rangle&~=~&\cos\phi_{ud}~|\eta_{ud}\rangle~-~\sin\phi_{s}~|\eta_{s}\rangle\,,\\
|\eta^\prime\rangle&~=~&\sin\phi_{ud}~|\eta_{ud}\rangle~+~\cos\phi_{s}~|\eta_{s}\rangle\,,
\end{alignat}
\end{subequations}
from which it follows that (\ref{etaTOeeBound}) and (\ref{etaprimeTOeeBound}) translate into relatively weak bounds on the axion isoscalar mixing angles: 
\begin{subequations}\label{etabounds}
\begin{alignat}{2}
|\theta_{a\eta}|&~=~|\cos\phi_{ud}~\thetaUD-\sin\phi_{s}~\thetaS| ~~\lsim~~\frac{0.014}{|\qe|} \,,\\
|\theta_{a\eta^{\prime}}|&~=~|\sin\phi_{ud}~\thetaUD+\cos\phi_{s}~\thetaS|~~\lsim~~\frac{0.01}{|\qe|}\,.
\end{alignat}
\end{subequations}
Taking $\phi_{ud}=39.8^\circ$ and $\phi_s=41.2^\circ$ from \cite{Escribano:2005qq} and conservatively assuming $|\qe|=1/2$ for concreteness, we display the bounds (\ref{etabounds}) in Fig.\,\ref{8Be4He}. 

In Fig.\,\ref{8Be4He} we also show contours of $\thetaUD$ and $\thetaS$ (dashed lines) for which the axionic contribution to $\eta^{(\prime)}\to e^+e^-$ becomes comparable to that of the SM. These contours can be interpreted as the sensitivity to $\thetaUD$ and $\thetaS$ in the hypothetical scenario of a future observation of these processes showing an $\OO(1)$ deviation from the branching ratios predicted in the SM, (\ref{etaTOeeSM}) and (\ref{etaprimeTOeeSM}). It goes without saying that the actual experimental sensitivity of future $\eta/\eta^\prime$-factories to $\thetaUD$ and $\thetaS$ could be substantially better if $\Br(\eta^{(\prime)}\to e^+e^-)$ could be measured with better than $\OO(10\%)$ precision, and if the uncertainties in the SM theoretical predictions could be reduced to the percent level.

\vspace{-0.2cm}
\subsection{Axio-hadronic $\eta$ and $\eta^\prime$ decays}
\label{secEtaHad}

Hadronic decay channels of $\eta$ and $\eta^\prime$ mesons could in principle be hiding promising signals of the QCD axion and/or other hadronically coupled ALPs. Amongst the most obvious modes are the three-body final states $\eta^{(\prime)}\to\pi^0\pi^0 a\,,\;\pi^+\pi^- a$, which have only recently been explored in the literature \cite{Aloni:2018vki,Landini:2019eck}. Indeed, the amplitudes for these processes receive a direct contribution from the leading order potential term in the chiral Lagrangian\footnote{More explicitly, these leading order quartic couplings do not contain derivatives of the axion field, nor do they ``descend'' from ordinary mesonic quartic terms via axion-meson mixing. Nonetheless, they are still consistent with the axion's pseudo-Goldstone nature because they are proportional to $\big(\sum_q m_q^{-1}\big)^{-1}$, and therefore vanish in the limit of a massless quark.}, and could in principle result in considerably large branching ratios. The difficulty with studying hadronic $\eta$ and $\eta^\prime$ decays lies in reliably predicting their rates. One of the earliest examples where this difficulty was encountered was in the calculation of $\eta\to 3\pi$, which was significantly underestimated by $\chiPT$ at leading order \cite{Cronin:1967jq,Osborn:1970nn,Gasser:1984pr,Kupsc:2007ce}. Indeed, it has long been understood that contributions from chiral logarithms and strong final state rescattering could not be neglected in the computation of $\eta\to 3\pi$ \cite{Kambor:1995yc,Borasoy:2005du,Anisovich:1996tx,Bijnens:2007pr,Guo:2015zqa,Guo:2016wsi,Colangelo:2018jxw}. Similarly, neither the total width nor the Dalitz phase space of $\eta^\prime\to\eta\pi\pi$ is properly described by $\chiPT$ at $\OO(p^2)$ \cite{HerreraSiklody:1999ss,Borasoy:2006uv}.

Previous studies \cite{Fariborz:1999gr,AbdelRehim:2002an} have shown that such intermediate-energy processes can be satisfactorily described by extending $\chiPT$ to include low-lying meson resonances carrying nonlinear realizations of $SU(3)_\chi$---such as vectors ($\rho$, $\omega$, $K^*$, $\phi$, ...), axial-vectors ($a_1$, $f_1$, $K_1$, ...), scalars ($a_0$, $f_0$, $\sigma$, $\kappa$, ...), and pseudoscalars ($\eta^\prime$, $\pi(1300)$, ...)---and assuming the principle of~~``resonance dominance'' (an extension of vector meson dominance), whereby the low-energy constants (LECs) of the $\OO (p^4)$ chiral Lagrangian are saturated by mesonic resonance exchange. This framework, dubbed \emph{Resonance Chiral Theory} ($\RchiT$), has been quite successful phenomenologically as an interpolating effective theory between the short-distance QCD description and the low-energy $\chiPT$ framework, by encoding the most prominent features of nonperturbative strong dynamics \cite{Ecker:1988te,Ecker:1989yg}. There does not appear to be consensus in the literature, however, on which low-lying resonances should be included as degrees-of-freedom in the $\RchiT$ Lagrangian, and which resonances should be regarded as dynamically generated poles due to strong S-wave interactions \cite{Oller:1997ti,Cirigliano:2003yq}. Examples of such ``ambiguous'' poles include the $\sigma(500)$ and the $\kappa(700)$.

In this subsection, we estimate the rates for $\eta\to\pi\pi a$ and $\eta^\prime\to\pi\pi a$ using $\RchiT$.
In both cases, we find that the leading order $\chiPT$ predictions for these decay rates are significantly modified by inclusion of resonance exchange amplitudes.
In particular, for $\eta\to\pi\pi a$, there is substantial destructive interference between the leading order amplitude from the $\OO(p^2)$ quartic term and the amplitudes generated by tree-level resonance exchange. This is corroborated by performing the same calculation in ordinary $\chiPT$ at $\OO(p^4)$, where one finds that the LECs, in particular $L_4$, $L_5$, and $L_6$, provide $\OO(1)$ contributions that destructively interfere with the $\OO(p^2)$ amplitude. For $\eta^\prime\to\pi\pi a$, the contributions from resonance exchange (alternatively, from $\chiPT$ interactions at $\OO(p^4)$) are the dominant effect in a significant portion of the parameter space, and may enhance this decay rate by an order of magnitude over the leading order $\chiPT$ prediction. The justification for favoring the $\RchiT$ framework over $\OO(p^4)$-$\chiPT$ for this calculation is that the former is expected to better capture the Dalitz phase-space of the final state, which is relevant when extracting the event acceptance due to momentum cuts in experimental analyses (in particular due to the $e^\pm$ selection criteria). Indeed, we will find that there is strong variation of the amplitude's momentum dependence as we vary the assumptions and parameters of the $\RchiT$ description, which implies a strong variation in the estimated sensitivity of existing and future experimental analyses. Unfortunately, the variations in these assumptions cannot be narrowed down without further input from experiment. Our main conclusion, therefore, is that one cannot reliably predict neither the total branching ratio, nor the Dalitz phase-space, of the decays $\eta^{(\prime)}\to\pi\pi a$. Under reasonable assumptions, our $\RchiT$-based estimates vary over two orders of magnitude in branching ratio, $\Br(\eta^{(\prime)}\to\pi\pi a)\sim \OO(10^{-4}-10^{-2})$. Nonetheless, this motivates dedicated reanalyses of existing data in final states of $\eta^{(\prime)}\to\pi\,\pi \,e^+ e^-$, as well as dedicated searches for $e^+ e^-$ resonances in these final states in future $\eta/\eta^{\prime}$ factories.

In order to motivate our use of $\RchiT$, and also to justify our later approximation of retaining only low-lying \emph{scalar} resonances, we begin by obtaining the main contributions to the amplitude $\mathcal{A}(\eta^{(\prime)}\to\pi^0\pi^0 a)$ in ordinary $\chiPT$ at $\OO(p^4)$. The Lagrangian is \cite{Gasser:1984gg}:
\begin{widetext}
\beqnarray\label{chiL}
\mathcal{L}_{\chiPT}\Big|_{\OO(p^4)}~=~&&\frac{f_\pi^2}{4}\,\Tr\big[D_\mu U^\dagger D^\mu U\big]~+~\frac{f_\pi^2}{4}\,\Tr\big[2B_0 M_q(a) U~+~\text{h.c.}\big]~-~\frac{1}{2}M_0^2\,\eta_0^2\\
+&&L_1\,\Tr\big[D_\mu U^\dagger D^\mu U\big]^2~+~L_2\,\Tr\big[D_\mu U^\dagger D_\nu U\big]\Tr\big[D^\mu U^\dagger D^\nu U\big]\nonumber\\
+&&L_3\,\Tr\big[D_\mu U^\dagger D^\mu U\,D_\nu U^\dagger D^\nu U\big]~+~L_4\,\Tr\big[D_\mu U^\dagger D^\mu U\big]\Tr\big[2 B_0 M_q(a) U+\text{h.c.}\big]\nonumber\\
+&&L_5\,\Tr\big[D_\mu U^\dagger D^\mu U\,(2 B_0 M_q(a) U+\text{h.c.})\big]~+~L_6\,\Tr\big[2 B_0 M_q(a) U+\text{h.c.}\big]^2\nonumber\\
+&&L_7\,\,\Tr\big[2 B_0 M_q(a) U-\text{h.c.}\big]^2~+~L_8\,\Tr\big[(2 B_0 M_q(a) U)\,(2 B_0 M_q(a) U)+\text{h.c.}\big]\nonumber\\
-&&i\,L_9\,\Tr\big[F_R^{\mu\nu}D_\mu U D_\nu U^\dagger+F_L^{\mu\nu}D_\mu U^\dagger D_\nu U\big]~+~L_{10}\,\Tr\big[U^\dagger F_R^{\mu\nu}\,U F_{L\,\mu\nu}\big]\,.\nonumber
\eeqnarray
\end{widetext}
Above, $f_\pi=92\;$MeV; $\delta^{ij}B_0=-\langle q^i{\bar{q}}^j\rangle/f_\pi^2\,$; $M_0$ parametrizes the $\OO$(GeV) contribution to the mass of the chiral singlet $\eta_0$ from the strong axial anomaly; $L_i$ ($i$=1,...,10) are the $\OO(p^4)$ $\chiPT$ low energy constants (LECs) (see, {\it e.g.}, \cite{Pich:2018ltt} for a review of $\chiPT$ and definitions of all the terms in (\ref{chiL})); $M_q(a)$ is the axion-dependent quark mass matrix, transforming as an octet spurion of $SU(3)_\chi$:

\beq\label{Mqa}
M_q(a)\equiv
\begin{pmatrix} 
\,m_u\,e^{i\,\qu a/f_a} & & \\
  & m_d\,e^{i\,\qd\,a/f_a}  &    \\
  &  & m_s~
\end{pmatrix},
\eeq
and $U$ is the non-linear representation of the pseudo-Goldstone chiral nonet:
\beqnarray
&&~~~~~~~~~~~~~~~~~~U~=~  \text{Exp}\bigg(i\,\frac{\sqrt{2}}{f_\pi}\varphi^a\lambda^a\bigg)\,\\
&&\text{with~~}\varphi^a\lambda^a \equiv\nonumber\\
&&\begin{pmatrix} 
\vspace{0.17cm}
\frac{\pi^0}{\sqrt{2}}+\frac{\eta_8}{\sqrt{6}}+\frac{\eta_0}{\sqrt{3}} & \pi^+ & K^+ \\
\vspace{0.17cm}
\pi^-  & -\frac{\pi^0}{\sqrt{2}}+\frac{\eta_8}{\sqrt{6}}+\frac{\eta_0}{\sqrt{3}}  &  K^0  \\
  K^- & \overline{K}^0 & -\frac{\eta_8}{\sqrt{3/2}}+\frac{\eta_0}{\sqrt{3}}~
\end{pmatrix}.\nonumber
\eeqnarray

Collecting the terms in (\ref{chiL}) that provide the dominant contributions to $\mathcal{A}(\eta^{(\prime)}\to\pi^0\pi^0 a)$, we obtain:
\begin{widetext}
\beqnarray\label{chiLquartic}
\!\!\!\mathcal{L}_{\chiPT}\Big|_{\OO(p^4)}~\supset&&~\frac{\sqrt{1+\hat{L}_4/2\;}}{1+\hat{L}_6}\,m_\pi^2\, f_\pi^2\,(\qu+\qd)\,\frac{m_um_d}{(m_u+m_d)^2}\;\Bigg[\widehat{\eta}_{ud}\,\widehat{\pi}^2\,\widehat{a}\\
&&-~\frac{\hat{L}_5}{2\,(1+\hat{L}_4/2)}\,\widehat\partial_\mu\widehat{\eta}_{ud}\;\widehat{\pi}\,\widehat\partial^{\,\mu}\widehat{\pi}\;\widehat{a}~-~\frac{(\hat{L}_5+2\hat{L}_4)}{4\,(1+\hat{L}_4/2)}\,\widehat{\eta}_{ud}\,\widehat\partial_\mu \widehat{\pi}\;\widehat\partial^{\,\mu}\widehat{\pi}\;\widehat{a}~+~\OO\bigg(\frac{m_\pi^2}{m_K^2}\bigg)\Bigg]\,,\nonumber
\eeqnarray
\end{widetext}
where we have defined the dimensionless fields
\beq
\widehat{\pi}\equiv\frac{\pi^0}{f_\pi}~,~~\widehat{\eta}_{ud}\equiv\left(\frac{1}{\sqrt{3}}\frac{\eta_8}{f_8}+\sqrt{\frac{2}{3}}\frac{\eta_0}{f_0}\right)~,~~ \widehat{a}\equiv\frac{a}{f_a}~,
\eeq
the dimensionless LECs:
\beq\label{Lhat}
\hat{L}_i~\equiv~\frac{32\,m_\eta^2}{f_\pi^2}\,L_i~\sim~\OO(10^3)\,L_i,
\eeq
and the dimensionless derivative $\widehat\partial_\mu\equiv \partial_\mu/m_\eta $.
The kinetic terms, omitted in (\ref{chiLquartic}), have been canonically normalized. While there is large variation in the literature of the inferred values for $L_4$ and $L_6$ from fits to experimental data, depending on assumptions and chosen observables, it is well established from fits to $f_K/f_\pi$ that $L_5$ is positive and relatively large, $L_5 \sim (1-3)\times 10^{-3}$. It is then easy to see from (\ref{chiLquartic}) and (\ref{Lhat}) that the contributions to $\mathcal{A}(\eta\to\pi^0\pi^0 a)$ from the first and second terms in (\ref{chiLquartic}) are comparable in magnitude and destructively interfere with each other. This leads to a suppressed rate for $\eta\to\pi^0\pi^0 a$ relative to the na\"{\i}ve $\OO(p^2)$ estimation in $\chiPT$, which, however, is quite sensitive to the value of $L_5$. On the other hand, the $\OO(p^2)$ contribution to $\mathcal{A}(\eta^\prime\to\pi^0\pi^0 a)$ from the first term in (\ref{chiLquartic}) may be subdominant to that of the second term, which is parametrically larger by a factor of $\OO(\widehat{L}_5\, m_\eta^{\prime\,2}/m_\eta^2)$. This may lead to an order-of-magnitude enhancement of the rate for $\eta^\prime\to\pi^0\pi^0 a$.

While it is now straightforward to extract $\chiPT$'s prediction for $\Br(\eta^{(\prime)}\to\pi^0\pi^0 a)$ using (\ref{chiLquartic}), we will instead pivot to $\RchiT$, from which ordinary $\chiPT$ can be recovered by integrating out the low-lying meson resonances. Under the assumption of resonance dominance, $\RchiT$ predicts that the relevant LECs contributing to $\eta^{(\prime)}\to\pi\pi a$ ($L_4$, $L_5$, and $L_6$) are saturated by the exchange of \emph{scalar} resonances. We will therefore omit the low-lying pseudoscalar, vector, and axial-vector resonances from our discussion. Following the notation in \cite{Ecker:1988te}, we have:

\beqnarray\label{RchiL}
\mathcal{L}_{\RchiT}~\supset~&&\frac{f_\pi^2}{4}\,\Tr\big[D_\mu U^\dagger D^\mu U\big]~-~\frac{1}{2}M_0^2\,\eta_0^2\\
&&+~\frac{f_\pi^2}{4}\,\Tr\big[2B_0 M_q(a) U~+~\text{h.c.}\big] \nonumber\\
&&+~c_d\,\Tr\big[S\,D_\mu U^\dagger D^\mu U\big] \nonumber\\
&&+~c_m\,\Tr\big[B_0\big(S\,M_q(a) \!+\! M_q(a)\,S\big)\, U~+~\text{h.c.}\big]\,,\nonumber
\eeqnarray
where $S$ is the low-lying $J^{PC}=0^{++}$ meson octet\footnote{Unlike some studies in the literature, we do not assume the large-$N_c$ limit and do not include a $0^{++}$ chiral singlet resonance in our analysis. Furthermore, following refs.\,\cite{Oller:1997ti,Cirigliano:2003yq}, we consider the broad $0^{++}$ state $f_0$(500) (a.k.a. $\sigma$(500)) a dynamically generated pole due to strong S-wave interactions, and therefore do not include it as a degree of freedom in (\ref{RchiL}),\,(\ref{RchiTScalarOctet}).},
\beq\label{RchiTScalarOctet}
S~=~
\begin{pmatrix} 
\frac{a_0}{\sqrt{2}}+\frac{f_0}{\sqrt{6}} & a_0^+ & * \\
a_0^-  & -\frac{a_0}{\sqrt{2}}+\frac{f_0}{\sqrt{6}} & *  \\
 *  & * & -\frac{f_0}{\sqrt{3/2}}~
\end{pmatrix}.
\eeq
\\
\\
Above, $a_0$ and $f_0$ are shorthand for $a_0(980)$ and $f_0(980)$, respectively \cite{Zyla:2020zbs}, and we have not explicitly identified the scalar mesons with nonzero strangeness, since they do not contribute to $\eta^{(\prime)}\to\pi\pi a$.
Accounting for the tadpole-induced nonzero vacuum expectation value of $f_0$,
\beq
\langle f_0 \rangle ~=~ -\,\frac{4\sqrt{2}}{\sqrt{3}}\,c_m\,\frac{\,(m_K^2-m_\pi^2/2)\,}{m_{f_0}^2}\,,
\eeq
and canonically normalizing the kinetic terms, we can extract from (\ref{RchiL}) the $\RchiT$ interactions contributing to $\eta^{(\prime)}\to\pi^0\pi^0 a$:
\begin{widetext}
\beqnarray\label{RchiLquartic}
\!\!\!\mathcal{L}_{\RchiT}~\supset&&~\frac{\sqrt{1+\frac{\widehat{c}_d\,\langle \widehat{f}_0\rangle}{\sqrt{3/2}\;}}}{1+\frac{\widehat{c}_m\,\langle \widehat{f}_0\rangle}{\sqrt{3/2}\;}}\,m_\pi^2\, f_\pi^2\,(\qu+\qd)\,\frac{m_um_d}{(m_u+m_d)^2}\\
&&\qquad\times~~\Bigg[\widehat{\eta}_{ud}\,\widehat{\pi}^2\,\widehat{a}~-~\frac{2\sqrt{2}}{\sqrt{1+\frac{\widehat{c}_d\,\langle \widehat{f}_0\rangle}{\sqrt{3/2}\;}}}\;\widehat{c}_m\,\widehat{a}_0\,\widehat{\pi}\,\widehat{a}~-~\frac{2\sqrt{2}}{\sqrt{3}}\;\widehat{c}_m\,\widehat{f}_0\,\widehat{\eta}_{ud}\,\widehat{a}\Bigg]\nonumber\\
&&+~\frac{m_\eta^2\, f_\pi^2}{\sqrt{1+\frac{\widehat{c}_d\,\langle \widehat{f}_0\rangle}{\sqrt{3/2}\;}}}\,\widehat{c}_d\;\Bigg[\sqrt{2}\;\widehat{a}_0\,\widehat{\partial}_\mu\widehat{\pi}\,\widehat{\partial}^{\,\mu}\widehat{\eta}_{ud}~+~\frac{1}{\sqrt{6}\,\sqrt{1+\frac{\widehat{c}_d\,\langle \widehat{f}_0\rangle}{\sqrt{3/2}\;}}}\;\widehat{f}_0\,\widehat{\partial}_\mu\widehat{\pi}\,\widehat{\partial}^{\,\mu}\widehat{\pi}\Bigg]\,,\nonumber
\eeqnarray
\end{widetext}
where, following the notation for the dimensionless fields and derivatives in (\ref{chiLquartic}), we have additionally introduced:
\beq
\widehat{a}_0~\equiv~\frac{a_0(980)}{f_\pi}~,\qquad\widehat{f}_0~\equiv~\frac{f_0(980)}{f_\pi}\,,
\eeq
and the dimensionless couplings:
\beq\label{cdhat}
\widehat{c}_d~\equiv~\frac{c_d}{\,(f_\pi/2)\,}~,\qquad\widehat{c}_m~\equiv~\frac{c_m}{\,(f_\pi/2)\,}\,.
\eeq

\vspace{0.5cm}

It is straightforward to recover (\ref{chiLquartic}) from (\ref{RchiLquartic}) by integrating out the scalar resonances and making the following identifications:
\beq
\qquad\qquad\qquad\qquad m_{a_0}~\approx~m_{f_0}~\approx~m_S\,,\qquad\qquad~~~
\eeq
\beq
L_4~=\;-\frac{c_d\,c_m}{3\,m_S^2}\,,~~L_5~=\;\frac{c_d\,c_m}{m_S^2}\,,~~L_6~=\;-\frac{c_m^2}{6\,m_S^2}\,.
\eeq

Early fits to $a_0(980)\to\pi\eta$ \cite{Ecker:1988te}, along with large-$N_c$ assumptions and imposition of short-distance constraints \cite{Pich:2002xy} (such as sum rules between two-point correlators of two scalar vs two pseudoscalar currents \cite{Golterman:1999au}, and vanishing of scalar form factors at $q^2\to\infty$ \cite{Jamin:2000wn,Jamin:2001zq}) have been used to estimate the scalar octet couplings to be $|c_d|\sim|c_m|\sim f_\pi/2$, with $c_d\,c_m>0$. Here, we conservatively vary these values by $\pm~20\%$ in our calculations of $\Gamma(\eta^{(\prime)}\to\pi\pi a)$, but retain, for simplicity, the assumption of $|c_d|=|c_m|$.

With the relevant interactions in (\ref{RchiLquartic}), we can then obtain the tree-level amplitude\footnote{We ignore corrections to $\mathcal{A}(\eta^{(\prime)}\to\pi\pi a)$ from $\pi\pi$ final state rescattering, based on the conclusions from \cite{Kambor:1995yc, Anisovich:1996tx} that these effects correct the $\eta\to3\pi$ amplitude by modest amounts of $\OO(10\%)$, and on Ref.~\cite{Borasoy:2005du}, which finds somewhat larger rescattering corrections, of $\sim 70\%$, which are still subdominant relative to other sources of uncertainties in our estimations.} $\mathcal{A}(\eta^{(\prime)}\to\pi\pi a)$ (see Fig.\,\ref{FeynmanDiagrams}):
\begin{widetext}
\begin{center}
\begin{figure}[t]
 \includegraphics[width=0.8\textwidth]{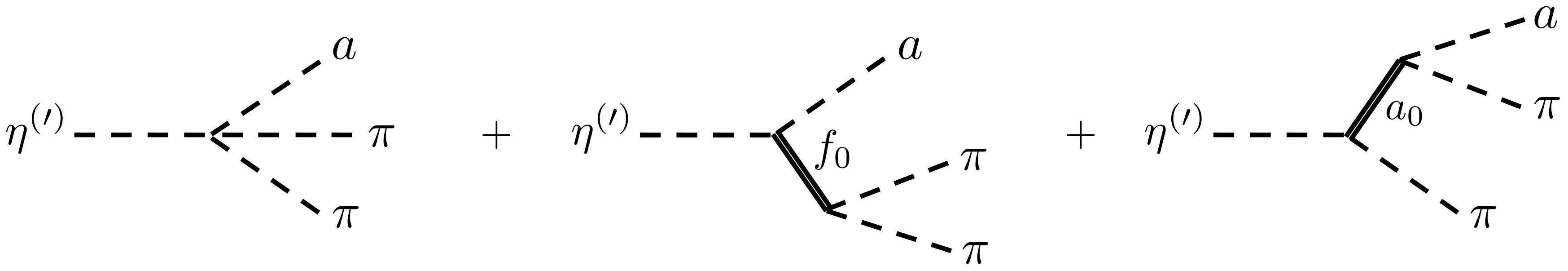}
\caption{\small Contributions to the amplitude $\mathcal{A}(\eta^{(\prime)}\to\pi\pi a)$ in the framework of $\RchiT$. Left graph: leading order quartic term. Middle and right graphs: exchange of low-lying scalar resonances.}
\label{FeynmanDiagrams}
\end{figure}
\end{center}
\beqnarray\label{etaAmplitude}
\mathcal{A}_{\eta^{(\prime)}\to\pi\pi a}~\equiv&&~\mathcal{A}(\eta^{(\prime)}\to\pi^0\pi^0 a)~~=~~\mathcal{A}(\eta^{(\prime)}\to\pi^+\pi^- a)\nonumber\\
=&&~2\,C_{\eta^{(\prime)}}\,\frac{f_\pi}{f_a}\,(\qu+\qd)\,\frac{m_\pi^2}{f_\pi^2}\,\frac{m_um_d}{(m_u+m_d)^2}\;\frac{\sqrt{1+\frac{\widehat{c}_d\,\langle \widehat{f}_0\rangle}{\sqrt{3/2}\;}}}{1+\frac{\widehat{c}_m\,\langle \widehat{f}_0\rangle}{\sqrt{3/2}\;}}\\
&&\times~~\Bigg[1~+~\frac{2\,\widehat{c}_d\,\widehat{c}_m}{1+\frac{\widehat{c}_d\,\langle \widehat{f}_0\rangle}{\sqrt{3/2}\;}}\,\bigg(\frac{1}{3}\,\frac{p_{\pi_1} .\, p_{\pi_2}}{~m_{f_0}^2\!-(p_{\pi_1}\!+p_{\pi_2})^2-i\,\Gamma_{f_0}m_{f_0}~}\nonumber\\
&&-~\frac{p_{\eta^{(\prime)}}  .\, p_{\pi_1}}{~m_{a_0}^2\!-(p_{\eta^{(\prime)}}\!-p_{\pi_1})^2-i\,\Gamma_{a_0}m_{a_0}~}~-~\frac{p_{\eta^{(\prime)}}  .\, p_{\pi_2}}{~m_{a_0}^2\!-(p_{\eta^{(\prime)}}\!-p_{\pi_2})^2-i\,\Gamma_{a_0}m_{a_0}~}\bigg)\Bigg]\,,\nonumber
\eeqnarray
\end{widetext}
where we have neglected subdominant contributions of $\OO(m_\pi^2/m^2_{\eta^{(\prime)}})$. Above, the equality between amplitudes with neutral vs charged pions is due to isospin symmetry; $p_{\eta^{(\prime)}}$, $p_{\pi_1}$, and $p_{\pi_2}$ are relativistic 4-momenta; and
\begin{subequations}\label{Ceta}
\begin{alignat}{2}
&C_\eta~&\equiv~~\frac{f_\pi}{f_8}\,\frac{\cos\theta_8}{\sqrt{3}}\,-\,\frac{f_\pi}{f_0}\,\frac{\sin\theta_0}{\sqrt{3/2}}\,,\\
&C_{\eta^\prime}\;&\equiv~~\frac{f_\pi}{f_8}\,\frac{\sin\theta_8}{\sqrt{3}}\,+\,\frac{f_\pi}{f_0}\,\frac{\cos\theta_0}{\sqrt{3/2}}\,.
\end{alignat}
\end{subequations}
For the $\eta/\eta^{\prime}$ mixing angles and decay constants above, we will adopt the values from the unconstrained fit in \cite{Escribano:2005qq}, namely, $\theta_8=-24^\circ$, $\theta_0=-2.5^\circ$, $f_8=1.51\,f_\pi$, and $f_0=1.29\,f_\pi$.

Finally, we can obtain the differential decay rate from (\ref{etaAmplitude}):
\beqnarray\label{diffDecayRate}
&&\!\!\!\!\!\! d\Gamma(\eta^{(\prime)}\to\pi\pi a)\\
&=&~\frac{1}{S_{\pi_1\pi_2}}\,\frac{(2\pi)^4}{2\,m_{\eta^{(\prime)}}}\,\left|\mathcal{A}_{\eta^{(\prime)}\to\pi\pi a}\right|^2\,d\Phi_3(p_{\eta^{(\prime)}};\,p_a,\,p_{\pi_1},\,p_{\pi_2})\nonumber\\
&=&~\frac{1}{S_{\pi_1\pi_2}}\,\frac{1}{(2\pi)^3}\frac{1}{32\,m^3_{\eta^{(\prime)}}}\left|\mathcal{A}_{\eta^{(\prime)}\to\pi\pi a}\right|^2\,dm_{\pi_1\pi_2}^2\,dm^2_{\pi_2 a}\,,\nonumber
\eeqnarray 
where $S_{\pi_1\pi_2}$ is the standard combinatorial factor ($S_{\pi^+\pi^-}\!=1$,  $S_{\pi^0\pi^0}\!=2!$\,), $d\Phi_3$ is the 3-body phase-space differential element, and, when obtaining the total decay rate, the integration over invariant masses $m_{\pi_1\pi_2}^2\!=\!(p_{\pi_1}\!+p_{\pi_2})^2$ and $m^2_{\pi_2 a}=(p_{\pi_2}\!+p_{a})^2\!=\!(p_{\eta^{(\prime)}}\!-p_{\pi_1})^2$ should be performed over the Dalitz phase-space (see, {\it e.g.}, the PDG review on {\it Kinematics} \cite{Zyla:2020zbs} for explicit expressions for the Dalitz plot boundaries).

\begin{figure}[t]
\centering
 \includegraphics[width=0.49\textwidth]{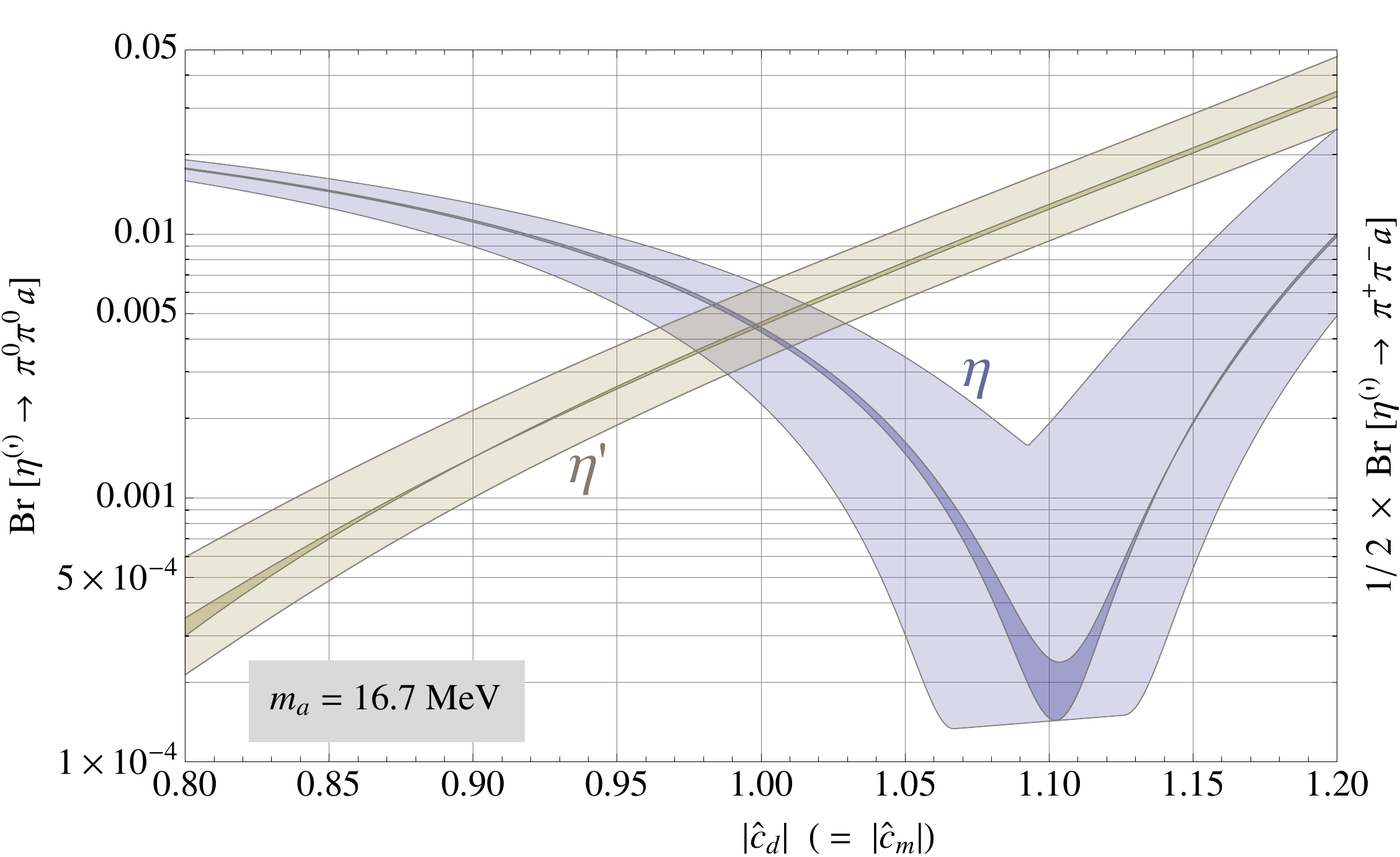}
\caption{\small Estimated branching ratios for $\eta^{(\prime)}\to\pi\,\pi\,a$ as a function of the scalar octet couplings to the light pseudoscalar mesons, {\it cf.} (\ref{RchiL}), (\ref{RchiLquartic}), and (\ref{cdhat}). The bands result from varying the masses and widths of the scalar resonances, $a_0$ and $f_0$, within their experimental uncertainties. For the dark narrow bands, their masses are fixed to $m_{a_0}=m_{f_0}=980\,$MeV, and their widths are varied within the ranges $\Gamma_{a_0}= (40-100)\,$MeV, $\Gamma_{f_0}= (10-200)\,$MeV. The broader bands result from additionally varying their masses within the ranges $m_{a_0},\,m_{f_0}= (960-1000)\,$MeV.}
\label{BrEta}
\end{figure}

In Fig.\,\ref{BrEta}, we show the branching ratios for $\eta^{(\prime)}\to\pi^0\pi^0 a$, $\eta^{(\prime)}\to\pi^+\pi^- a$ (computed by integrating the differential decay rate in (\ref{diffDecayRate}) over the final state phase space) as a function of the $\RchiT$ couplings $\widehat{c}_d$, $\widehat{c}_m$ of the low-lying scalar octet to the pseudo-Goldstone mesons; see eqs.\,(\ref{RchiL}), (\ref{RchiLquartic}), and (\ref{cdhat}). As mentioned previously, we assume, for simplicity, that $\widehat{c}_d=\widehat{c}_m$, and vary their magnitudes by $\pm20\%$ around their expected values of $|\widehat{c}_d|=|\widehat{c}_m|=1$ in the large-$N_c$ limit. The range covered by the bands are due to the uncertainties in the masses and widths of the scalar resonances $a_0(980)$ and $f_0(980)$---following the PDG \cite{Zyla:2020zbs}, we varied these parameters independently within the following ranges: $m_{a_0},\,m_{f_0}=(960-1000)$~MeV, $\Gamma_{a_0}=(40-100)$~MeV, $\Gamma_{f_0}=(10-200)$~MeV.

The lack of predictive power of our treatment, with an estimated range of branching ratios spanning two orders of magnitude, $\Br(\eta^{(\prime)}\to\pi\pi a)\sim \OO(10^{-4}-10^{-2})$, is due to the $\chiPT$ and $\RchiT$ parameters falling on a special range of values that, within uncertainties, can lead to substantial destructive inference between the LO amplitude and the amplitudes originating from  exchange of low-lying scalar resonances. This is perhaps unsurprising, considering that even the SM hadronic decays of the $\eta$ and $\eta^\prime$ could not be correctly predicted,
but only ``postdicted,'' and their experimentally determined branching ratios and Dalitz plot parameters have been used to verify the validity of various treatments and assumptions, such as $\RchiT$, QCD sum rules, large-$N_c$ limit, dispersive methods, etc \cite{Kambor:1995yc,Borasoy:2005du,Anisovich:1996tx,
Bijnens:2007pr,Guo:2015zqa,Guo:2016wsi,Colangelo:2018jxw,HerreraSiklody:1999ss,Borasoy:2006uv,Fariborz:1999gr,AbdelRehim:2002an}.

In particular, the upper range of our estimations, $\Br(\eta^{(\prime)}\to\pi\pi a)\sim \OO(10^{-2})$, is probably excluded or in tension with observations, though no dedicated searches for an $e^+e^-$ resonance in $\eta^{(\prime)}\to\pi\,\pi\, e^+e^-$ final states have ever been performed, to the best of our knowledge. However, the lower range $\Br(\eta^{(\prime)}\to\pi\pi a)\sim \OO(10^{-4}-10^{-3})$ likely remains experimentally allowed, and within the sensitivity of upcoming $\eta/\eta^\prime$ factories, such as the JLab Eta Factory (JEF) and the REDTOP experiment.

The most recent and precise measurement of the SM decay $\eta\to\pi^+\pi^- (\gamma^*\to e^+ e^-)$, which shares the same final state of $\eta\to\pi^+\pi^- a$, was performed by the KLOE Collaboration at the Frascati $\phi$-factory DA$\Phi$NE \cite{Ambrosino:2008cp}. While their measurement yielded $\Br(\eta\to\pi^+\pi^-e^+ e^-)=(2.68\pm0.09_\text{stat}\pm0.07_\text{syst})\times10^{-4}$, it is nontrivial to infer any bounds from this analysis on $\Br(\eta\to\pi^+\pi^- a)$. This is because, without proper Monte Carlo simulations, one cannot determine how the background rejection requirements would have affected the $\eta\to\pi^+\pi^- a$ signal efficiency. In particular, this search rejected events with $m_{e^+e^-}<15$\;MeV whose reconstructed $e^+e^-$ vertex was within a 2.5\,cm distance from the beampipe. This cut could have significantly impacted the acceptance of the axion signal, depending on the $m_{e^+e^-}$ experimental resolution. Other event selection requirements on the momenta of the $\pi^\pm$ and $e^\pm$ charged tracks could in principle have rejected a large fraction of the axion signal as well.

An earlier measurement of $\Br(\eta\to\pi^+\pi^- (\gamma^*\to e^+ e^-))$  by the CELSIUS/WASA Collaboration observed, in hindsight, an upward fluctuation of the expected signal \cite{Bargholtz:2006gz,Berlowski:2007aa}. Indeed, considering KLOE's more precise measurement of this branching ratio, the CELSIUS/WASA analysis should have expected 10 SM signal events. It observed 24 events in the signal region, of which it determined that 7.7 were from background, and 16.3 were from the SM signal. Assuming, conservatively, that the 14 ``excess events'' were instead due to $\eta\to\pi^+\pi^- a$ decays, and taking into account the relative signal acceptance due to the minimum transverse momentum requirement of $|\vec{p}_T|>20$\;MeV for charged particles\footnote{We performed a simple MC event simulation to estimate the geometric acceptance resulting from the event selection requirement of $|\vec{p}_T^{\;e^\pm}|>20$ MeV, properly taking the momentum dependence of the amplitudes into account. This was done for both the SM signal using the amplitude in \cite{Bargholtz:2006gz}, as well as for the axion signal, assuming a few $\RchiT$ benchmark parameters (see discussion below). We neglected contributions to the signal efficiency from other event selection requirements, and worked in the approximation of $\eta$ mesons decaying at rest.}, we estimate that branching ratios as large as $\Br(\eta\to\pi^+\pi^- a)\sim(1-3)\times 10^{-3}$ could be compatible with the CELSIUS/WASA measurement, although, without access to non-public information on details of the experimental analysis, this estimation is at the level of an educated guess.

Finally, the two existing measurements of $\Br(\eta^\prime\to\pi^+\pi^- (\gamma^*\to e^+ e^-))$, performed independently by the CLEO \cite{Naik:2008aa} and BESIII \cite{Ablikim:2013wfg} Collaborations, were combined by the PDG \cite{Zyla:2020zbs} to give $\Br(\eta^\prime\to\pi^+\pi^- e^+ e^-)=\big(2.4^{\,+\,1.3}_{\,-\,0.9}\big)\times 10^{-3}$. However, both experimental analyses reported large external photon-conversion backgrounds in the signal region, peaked in the range $m_{e^+e^-}=(8-25)$\;MeV (CLEO; see Fig.\,2(d) of \cite{Naik:2008aa}) and $m_{e^+e^-}=(10-20)$\;MeV (BESIII; see Fig.\,2 of \cite{Ablikim:2013wfg}). Events falling within these $m_{e^+e^-}$ windows were excluded from the analyses' inference of the SM branching ratio. Since events from $\eta^\prime\to\pi^+\pi^- a$ would have fallen precisely in this region where the photon conversion background peaked, it is difficult to estimate how strong a potential axion signal could have been. Simply requiring that the axion signal strength does not overpredict the number of events attributed to photon-conversion yields a conservative limit of $\Br(\eta^\prime\to\pi^+\pi^- a)\lsim\text{few}\times 10^{-2}$, which is not particularly useful.

\begin{figure}[b]
\centering
 \includegraphics[width=0.48\textwidth]{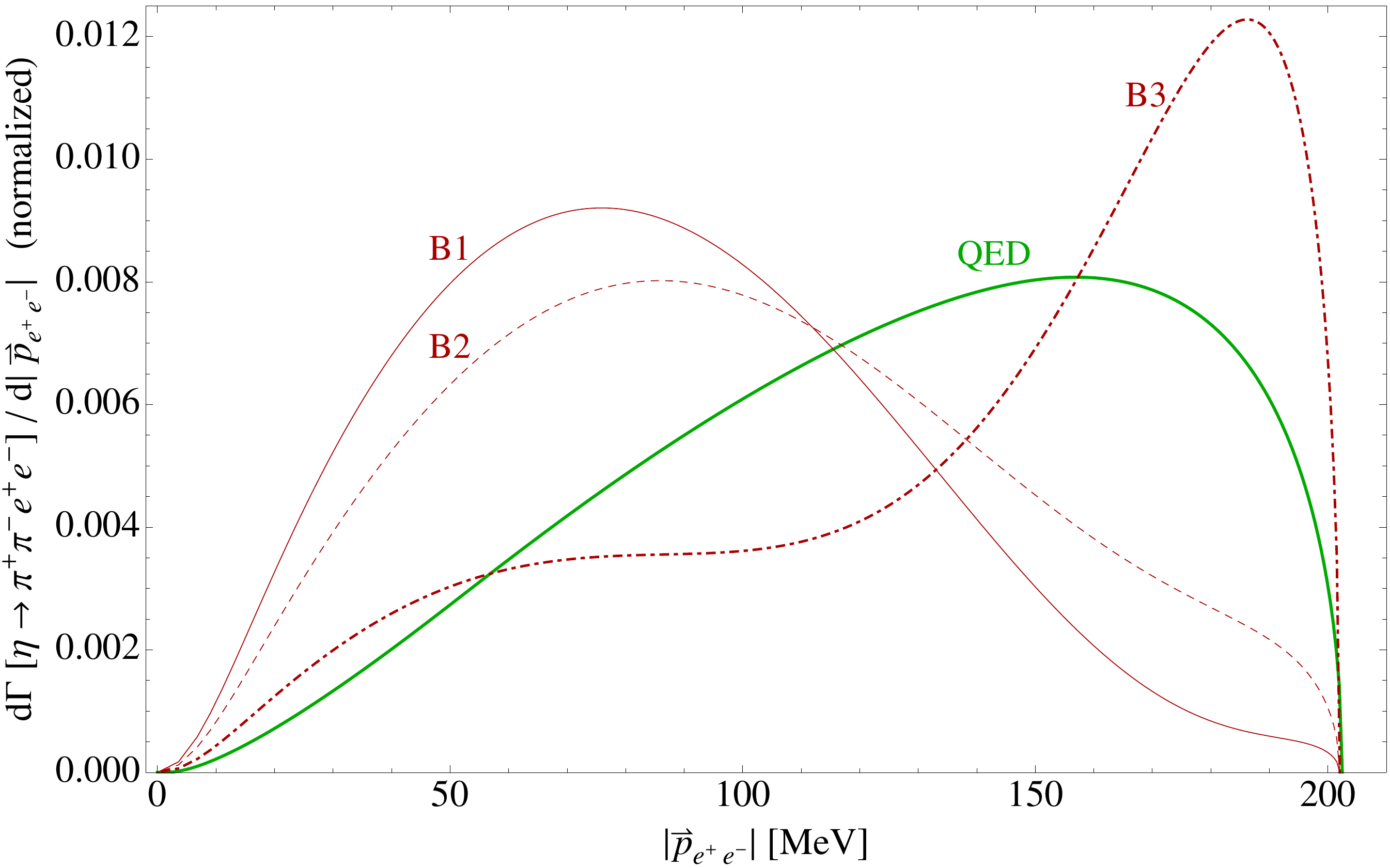}
\caption{\small The differential rate for $\eta\to\pi^+\pi^- a$ as a function of $|\vec{p}_{e^+e^-}|\equiv |\vec{p}_{e^+}+\vec{p}_{e^-}| = \vec{p}_a$, for three benchmark choices of $\RchiT$ parameters specified in Table\,\ref{BenchmarkTable}. For comparison, we also show the differential rate of the SM process $\eta\to\pi^+\pi^- e^+ e^-$, labeled ``QED''.}
\label{PhaseSpaceBenchmarks}
\end{figure}

We end this section by remarking that an additional challenge with estimating the sensitivity of current and future experiments to $\eta^{(\prime)}\to\pi\pi a$ is the uncertainty in the final state Dalitz phase-space, which affects the signal acceptance resulting from event selection cuts. Consider, for instance, the differential decay rate $d\Gamma(\eta\to\pi^+\pi^- a)/d|\vec{p}_a|$ as a function of the axion's 3-momentum $|\vec{p}_a|$. The dependence of this rate on $|\vec{p}_a|=|\vec{p}_{e^+}+\vec{p}_{e^-}|\equiv |\vec{p}_{e^+e^-}|$ varies dramatically depending of the numerical values chosen for the masses, widths, and couplings of the scalar resonances $a_0$ and $f_0$. We illustrate this effect in Fig.\,\ref{PhaseSpaceBenchmarks}, where we plot the differential decay rate $d\Gamma(\eta\to\pi^+\pi^- e^+e^-)/d|\vec{p}_{e^+e^-}|$ as a function of $|\vec{p}_{e^+e^-}|$ for three different $\RchiT$ benchmarks---corresponding to different choices of masses and widths for $a_0$ and $f_0$ within uncertainties (see Table\,\ref{BenchmarkTable})---as well as for the SM decay $\eta\to\pi^+\pi^- (\gamma^*\to e^+ e^-)$ \cite{Landsberg:1986fd} (labeled as ``QED'' in Fig.\,\ref{PhaseSpaceBenchmarks}).
While these different benchmark points yield close predictions for $\Br(\eta\to\pi^+\pi^- a)$, their predictions for $d\Gamma(\eta\to\pi^+\pi^- a)/d|\vec{p}_{a}|$ differ dramatically, as shown in Fig.\,\ref{PhaseSpaceBenchmarks}. In particular, for high enough cuts on the charged lepton momenta $\vec{p}_{e^\pm}$, the signal acceptance of benchmark B1 could be significantly lower than that of B3 (and of the SM decay $\eta\to\pi^+\pi^- \gamma^*$). Indeed, this could lead to a variation of as much as an order of magnitude in the expected sensitivity of experimental searches.

\begin{widetext}
\begin{center}
\begin{table}[h]
\begin{tabular}{|l||c|c|c|c|c||l|}
\hline
&$m_{a_0}$ [MeV] & $\Gamma_{a_0}$ [MeV] & $m_{f_0}$ [MeV] & $\Gamma_{f_0}$ [MeV] & $|\widehat{c}_d| = |\widehat{c}_m|$  & $\Br(\eta\to\pi^+\pi^- a)$ \\
\hline
B1 &980 & 40 & 980 & 200 & 1.125 & $~~0.96\times10^{-3}$\\
B2 &980 & 50 & 980 & 100 & 1.125 & $~~~\,1.1\times10^{-3}$\\
B3 &1000 & 50 & 1000 & 100 & 1.125 & $~~0.49\times10^{-3}$\\
\hline
\end{tabular}
\caption{\label{BenchmarkTable} Benchmarked $\RchiT$ parameters for the examples in Fig.\,\ref{PhaseSpaceBenchmarks}, and the resulting prediction for the total decay rate of $\eta\to \pi^+\pi^- a$ .}
\label{BenchmarkTable}
\end{table}
\end{center}
\end{widetext}

\section{Kaon decays}
\label{secKaon}

We conclude our study by exploring signals of the \emph{piophobic} QCD axion in rare Kaon decays. Although the main focus of ongoing and near-future rare Kaon decay experiments---such as NA62 at CERN \cite{NA62:2017rwk} and KOTO at J-PARC \cite{Iwai:2012qya}---has been on $K\to\pi\nu\bar{\nu}$, there is an under explored opportunity to search for BSM resonances in $e^+e^-$ final states with low $m_{e^+e^-}$, motivated not only by the \emph{piophobic} QCD axion, but also by visibly decaying ALPs and dark photons more generally \cite{Gori:2020xvq}. Furthermore, the highly suppressed $a\to\gamma\gamma$ decay mode might be a competitive final state in Kaon decay searches for which $\gamma\gamma$ backgrounds in the $m_{\gamma\gamma}\sim 17\;\text{MeV}$ signal region are tamer than the $e^+e^-$ backgrounds. In such cases, final states with $K\to(a\to\gamma\gamma)+\text{SM}$ can be obtained by combining the relevant branching ratios $\Br(K\to a+\text{SM})$ estimated in this section with $\Br(a\to\gamma\gamma)$ in (\ref{BRaTogg}).

The appearance of the axion in Kaon decay final states occurs via mixing with the neutral octet mesons, $\pi^0$, $\eta$, and $\eta^\prime$. Therefore, the axionic amplitudes can be obtained from ordinary SM amplitudes properly reweighted by axion-meson mixing angles. While this prescription is straightforward for estimating ``\emph{axio-leptonic}'' Kaon decays such as $K^+\to\mu^+\nu_\mu\, a$ (as we will show in Subsec.\,\ref{secKpDecays}), it is ambiguous for ``\emph{axio-hadronic}'' Kaon decays such as $K^+\to\pi^+ a$, $K_{S,\,L}^0\to\pi^0 a$, and $K_{L}^0\to\pi\pi a$.
Firstly, the two-body hadronic width of the CP-even neutral Kaon, $\Gamma(K_S^0\to\pi^0\pi^0,\,\pi^+\pi^-)\approx0.73\times 10^{-5}\;\text{eV}$, is enhanced by roughly three orders of magnitude relative to the two-body hadronic width of the charged Kaon, $\Gamma(K^+\to\pi^+\pi^0)\approx1.1\times10^{-8}\;\text{eV}$.
In $\chiPT$, this enhancement is parametrized as a large disparity in the magnitudes of the Wilson coefficients of the possible $\Delta S =1$ operators \cite{Cronin:1967jq,Bernard:1985wf,Kambor:1989tz,Cirigliano:2011ny}. Specifically, the coefficient of an $SU(3)_\chi$-octet ($\Delta I=1/2$) operator is larger than the coefficient of the leading order 27-plet ($\Delta I=3/2$) operator by a factor of $\sim30$. Secondly, phenomenologically, there are at least two choices of $\Delta S =1$ octet operators that could be responsible for this so-called ``\emph{octet enhancement}'' ({\it a.k.a.} ``$\Delta I =1/2$ \emph{enhancement}'') in Kaon decays \cite{Gerard:2000jj,Crewther:2013vea,Buras:2014maa}, namely,

\begin{subequations}\label{O8O8prime}
\begin{alignat}{1}
&O_8^{\,(\Delta S =1)\,}~=~~~g_8\,f_\pi^2\,\Tr\big(\lambda_{ds}\, \partial_\mu U\, \partial^\mu U^\dagger\big)\,+\,\text{h.c.},\label{O8}\\
&\nonumber\\
&O_8^{\prime\,(\Delta S =1)}~=\nonumber\\
&-g_{8}^\prime\;\frac{f_\pi^2}{\Lambda^2}\;\Tr\big(\lambda_{ds}\,2B_0M^\dagger_q(a)\,U^\dagger\big)\;\Tr\big(\partial_\mu U\, \partial^\mu U^\dagger\big)\;+\;\text{h.c.}\,,\label{O8prime}
\end{alignat}
\end{subequations}
where $\lambda_{ds}\equiv(\lambda_6+i\lambda_7)/2$,
$\Lambda\sim 2m_K$ is a natural $\chiPT$ cutoff, and the operators above occur at different orders in the chiral expansion: $O_8$ at $\OO(p^2)$, and $O_8^\prime$ at $\OO(p^4)$. Fitting existing data on $K\to\pi\pi$ and $K\to\pi\pi\pi$, while treating the coefficients $g_8$ and $g_8^\prime$ in (\ref{O8O8prime}) on equal footing, yields $|g_8+g_8^{\prime}|\simeq 0.78\times 10^{-7}$ \cite{Kambor:1991ah,Bijnens:2002vr}. In order to break this degeneracy in the fit, one must invoke
the standard assumption under na\"{\i}ve power counting that $\Delta S =1$ \emph{octet enhancement} should appear at lowest order in the chiral expansion, and therefore, $g_8^\prime\ll g_8~\Rightarrow~|g_8|\simeq 0.78\times 10^{-7}$. However, there is no first principles derivation of this choice, and it could be incorrect. For example, in the Resonance Chiral Theory framework discussed in the previous section, it is easy to speculate that the origin of $\Delta S =1$ octet enhancement could be due to the weak interactions inducing a mixing between $K_S^0$ and a broad $J^{PC}=0^{++}$ resonance, such as the $\sigma(500)$\footnote{Indeed, a naive dimensional analysis estimation of $g_8^\prime\sim |G_F\sin\theta_{c\,} f_\pi^{2\,} \Lambda^2/m_{0^{++}}^2|\sim\OO(10^{-7})$ does not immediately rule out this hypothesis. It is unclear whether a Dalitz plot analysis of $K_L^0\to3\,\pi$ data could distinguish it from the alternative description of octet enhancement.}. Upon integration of the low-lying resonances, this effect would be captured by the operator $O_8^\prime$ in (\ref{O8prime}), leading instead to $g_8\ll g_8^\prime~\Rightarrow~|g_8^\prime|\simeq 0.78\times 10^{-7}$.

This ambiguity
directly affects predictions for rare Kaon decays to the \emph{piophobic} axion, since the $\Delta S=1$ octet operators $O_8$ and $O_8^\prime$ contribute differently to the amplitudes $\mathcal{A}(K^+\to\pi^+ a)$, $\mathcal{A}(K_{S,\,L}^0\to\pi^0 a)$, and $\mathcal{A}(K_{L}^0\to\pi\pi a)$. In what follows, we will estimate the rates for various axionic Kaon decays in both scenarios, $g_8\gg g_8^\prime$ and $g_8\ll g_8^\prime$. We will show that in the case of $g_8\gg g_8^\prime$, all amplitudes $\mathcal{A}(K^+\to\pi^+ a)$, $\mathcal{A}(K_{S,\,L}^0\to\pi^0 a)$, and $\mathcal{A}(K_{L}^0\to\pi\pi a)$ are octet-enhanced, leading to higher axionic Kaon decay rates, and when relevant, more stringent constraints on the mixing angles $\thetaUD$ and $\thetaS$. Conversely, in the scenario with $g_8\ll g_8^\prime$, the rates $\Gamma(K^+\to\pi^+ a)$ and $\Gamma(K_{S,\,L}^0\to\pi^0 a)$ are significantly reduced, relaxing the otherwise strong constraints on $\thetaUD$ and offering an exciting prospect for searching for these signals in near-future rare Kaon decay experiments.

In upcoming subsections, we will normalize the calculated axio-hadronic rates to analogous Kaon decay rates in the SM. For later reference, we quote here the dependence of the relevant SM Kaon decay amplitudes on $g_8$, $g_8^\prime$, as well as $g_{27}$, the coefficient of the $\Delta S =1$ 27-plet operator at $\OO(p^2)$ in $\chiPT$, which is given by:
\beq\label{O27}
O_{27}^{\,(\Delta S =1)\,}~=~~g_{27}\;f_\pi^2\;T^{ij}_{kl}\;\big(U^\dagger\partial_\mu U\big)^k_{\;i}\,\big(U^\dagger\partial^\mu U\big)^l_{\;j}\,.
\eeq
Above, $T^{ij}_{kl}$ are Clebsch-Gordan coefficients that project the 27-plet, $\Delta I=3/2$ part of the interaction \cite{Kambor:1989tz}. The contributions from (\ref{O8}), (\ref{O8prime}), and (\ref{O27}) to the two-body hadronic Kaon decays of interest are \cite{Kambor:1991ah,Donoghue:1992dd}:
\beqnarray
\mathcal{A}(K^0_S\to\pi^+\pi^-)~&=&~2\,(g_8+g_8^\prime+g_{27})\frac{m_K^2}{f_\pi}\nonumber\\
&&-\,2\,(g_8+2\,g_8^\prime+g_{27})\frac{m_\pi^2}{f_\pi}   \,,\label{KsToPiPi}\\
&&\nonumber\\
\mathcal{A}(K^+\to\pi^+\pi^0)~&=&~3\,g_{27}\frac{\big(m_K^2-m_\pi^2\big)}{f_\pi}\,.\label{KpToPiPi}
\eeqnarray

As alluded to earlier, the hadronic $K^+$ decay amplitude is not octet-enhanced, and its consequent narrow width relative to that of $K^0_S$ is parametrized by a hierarchy between the 27-plet and octet coefficients:
\beq\label{g27}
g_{27}~\simeq~2.5\times10^{-9}~\approx~0.032\times\,|g_8+g_8^{\prime}|.
\eeq

Finally, for the relevant hadronic three-body decay of $K_L^0$, we have \cite{Kambor:1991ah,Donoghue:1992dd}:
\beqnarray
\!\!\!\!\mathcal{A}(K^0_L\to\pi^+\pi^-\pi^0)\,=\,\frac{(g_8+g_8^\prime+2g_{27})}{3}\frac{m_K^2}{f_\pi^2}&&\nonumber\\
-\,g_8^\prime\,\frac{m_\pi^2}{f_\pi^2}\,+\,(g_8+g_8^\prime-\frac{5}{2}g_{27})\frac{m_\pi^2\,Y}{f_\pi^2}&&\,,\label{AmpKLongSM}
\eeqnarray
where $Y$ is one of the standard Dalitz plot variables, defined as:
\begin{subequations}\label{DalitzY}
\begin{alignat}{1}
&Y~\equiv~(s_3-s_0)/m_\pi^2\,,\\
&s_i~\equiv~(p_K-p_i)^2\Big|_{i=1,\,2,\,3}\,,\\
&s_0~\equiv~\frac{(s_1+s_2+s_3)}{3}\,,
\end{alignat}
\end{subequations}
with $p_1$ and $p_2$ referring to the four-momenta of the charged pions, and $p_3$ the four-momentum of the neutral daughter particle \footnote{Note that in Subsec.\,\ref{secKLdecays3body}, $p_3$ will refer to the axion's four-momentum.}, in this case $\pi^0$.

\subsection{$K^+$ decays}
\label{secKpDecays}
\subsubsection{Axio-leptonic $K^+$ decays}

The amplitude for the \emph{axio-leptonic} decay $K^+\to\ell^+\nu_\ell\,a$ can be easily related to the SM semi-leptonic amplitudes via the Ademollo-Gatto theorem \cite{Ademollo:1964sr}, which states that the matrix elements of flavor-changing electroweak current operators can only deviate from their $SU(3)_\chi$-symmetric values to second order in chiral symmetry breaking \cite{Leutwyler:1984je,Lebed:1991sq}.

This implies that, at zero momentum transfer, the following $SU(3)_\chi$ relations hold:

\begin{subequations}
\begin{alignat}{1}
\big\langle\eta_8\,\big|\,\bar{s}\gamma^\mu u\,\big|\,K^+\big\rangle\Big|_{q^2=0}~&=~\sqrt{3}\;\big\langle\pi^0\,\big|\,\bar{s}\gamma^\mu u\,\big|\,K^+\big\rangle\Big|_{q^2=0}\nonumber\\
&~+~\OO(\epsilon^2)\,, \\
\big\langle\eta_0\,\big|\,\bar{s}\gamma^\mu u\,\big|\,K^+\big\rangle\Big|_{q^2=0}~&=~~\OO(\epsilon^2)\,,
\end{alignat}
\end{subequations}
where $\epsilon$ is a measure of $SU(3)_\chi$ breaking. Then, since $|a\rangle=\theta_{a\pi}|\pi^0\rangle+\theta_{a\eta_8}|\eta_8\rangle+\theta_{a\eta_0}|\eta_0\rangle$, we have:

\beqnarray\label{axiolepAmp}
&&\big\langle a\,\big|\,\bar{s}\gamma^\mu u\,\big|\,K^+\big\rangle\Big|_{q^2=0}\\
&&=\big(\theta_{a\pi}+\sqrt{3}\,\theta_{a\eta_8}\big)\big\langle\pi^0\,\big|\,\bar{s}\gamma^\mu u\,\big|\,K^+\big\rangle\Big|_{q^2=0}+\OO(\epsilon^2)\,,\nonumber\\
&&=\big(\theta_{a\pi}+\,\thetaUD\!-\sqrt{2}\,\thetaS\big)\big\langle\pi^0\,\big|\,\bar{s}\gamma^\mu u\,\big|\,K^+\big\rangle\Big|_{q^2=0}+\OO(\epsilon^2)\,.\nonumber
\eeqnarray
Neglecting the difference in phase space, as well as finite momentum-transfer and $SU(3)_\chi$ breaking corrections, which amount to $\OO(10\%)$ \cite{Donoghue:1992dd}, it then follows from (\ref{axiolepAmp}) that:
\\
\\
\beqnarray
&&\Br(K^+\to\ell^+\nu_\ell\,a)~\approx\label{axiolepBr}\\
&&~~~~\big|\theta_{a\pi}+\,\thetaUD\!-\sqrt{2}\,\thetaS\big|^2~\Br(K^+\to\ell^+\nu_\ell\,\pi^0)\,,\nonumber
\eeqnarray
In the specific case of a muonic final state, (\ref{axiolepBr}) yields:
\beqnarray
&&\Br(K^+\to\mu^+\nu_\mu\,a)~\approx\\
&&~~~~0.84\times10^{-8}~\left|\frac{\,\theta_{a\pi}+\,\thetaUD\!-\sqrt{2}\,\thetaS\,}{5\times10^{-4}}\right|^2\,.\nonumber
\eeqnarray
%
%
In Fig.\,\ref{KaonPlots}, we show the hypothetical reach of $K^+\to\mu^+\nu_\mu\,a$ to the axion isoscalar mixing angles, assuming an experimental sensitivity to branching ratios $\Br(K^+\to\mu^+\nu_\mu\,a)\gsim10^{-8}$. Note that this branching ratio sensitivity figure has been chosen to facilitate comparison between different axionic Kaon decay modes (to be discussed in upcoming subsections), and is not informed by any experimental sensitivity projections.

\subsubsection{Axio-hadronic $K^+$ decays}
\label{secKpPia}

For axio-hadronic Kaon decays, we must first obtain the contributions from operators (\ref{O8}), (\ref{O8prime}), and (\ref{O27}) to the amplitudes for $K^+\to\pi^+\varphi^{*}$  ($\varphi=\pi^{0},\,\eta_{ud},\,\eta_{s}$). Putting $K^+$ and $\pi^+$ on shell, we have:
\vspace{-0.5cm}
\begin{widetext}
\begin{center}
\begin{figure}[t]
 \includegraphics[width=1\textwidth]{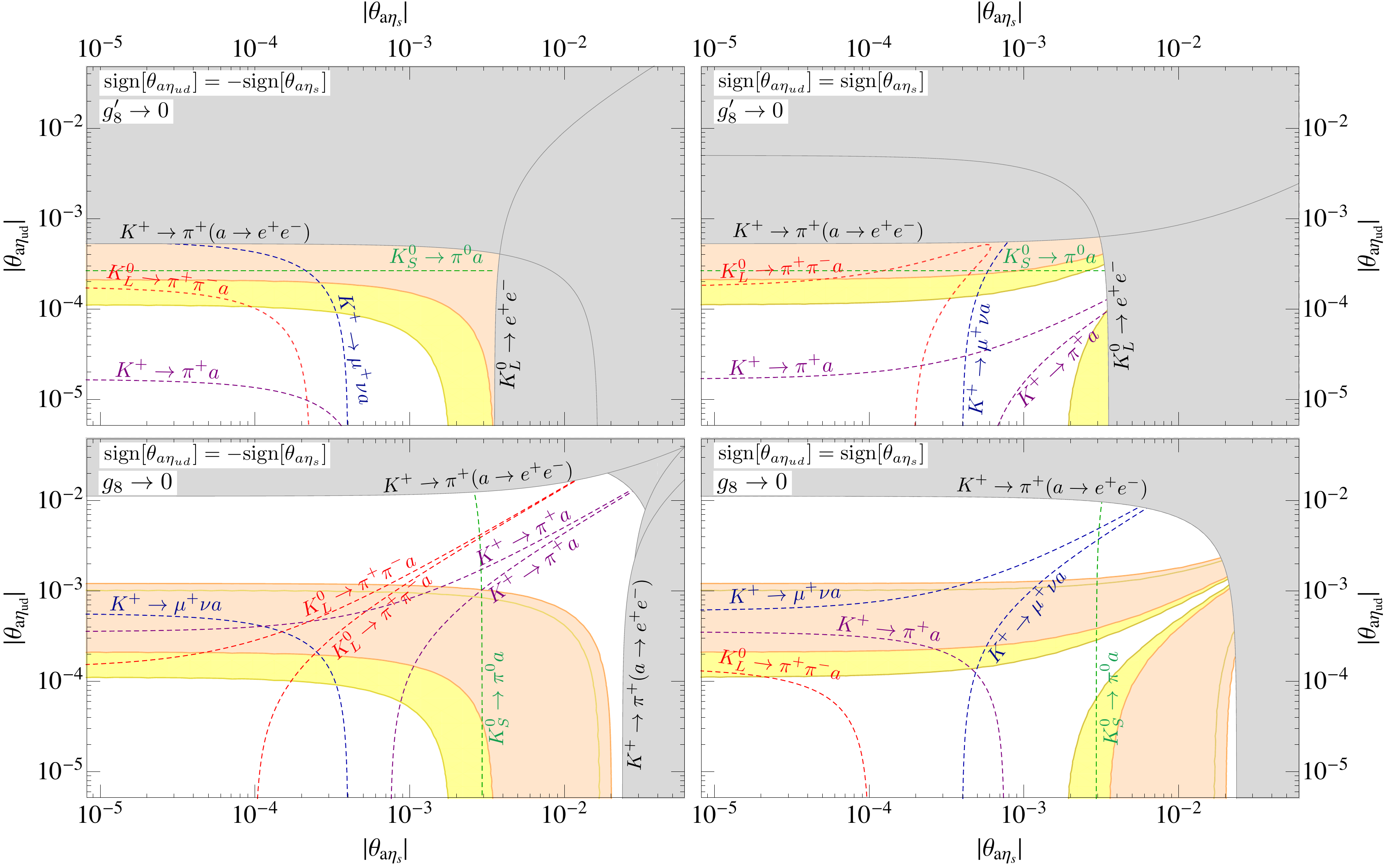}
\caption{\small \emph{(Disclaimer: not intended as a realistic experimental sensitivity projection.)} The dashed lines show the reach of various axionic Kaon decays modes in the parameter space of the axion isoscalar couplings, assuming a common branching ratio sensitivity benchmark of $10^{-8}$ for all decay channels. The upper (lower) plots assume that octet enhancement in $\chiPT$ is realized through operator $O_8$ ($O_8^\prime$) defined in \ref{O8} (\ref{O8prime}). The left (right) plots assume opposite (same) relative sign between $\thetaUD$ and $\thetaS$. The orange and yellow bands favored by the $\Be$ and $\He$ anomalies are the same as in Fig.\,\ref{8Be4He}. The shaded gray regions are excluded by the conservative upper bound $\Br(K^+\to \pi^+ (a\to e^+e^-))\lsim 10^{-5}$, and by the observed rate for $K_L^0\to e^+ e^-$, {\it cf.} \ref{KLeeExp}.}
\label{KaonPlots}
\end{figure}
\end{center}
\begin{subequations}
\begin{alignat}{1}
\mathcal{A}\big(K^+\to\pi^+{\pi^{0}}^{*}\big)~&=~3\,g_{27}\,\frac{m_K^2}{f_\pi}\,-\,(g_8+4g_{27})\,\frac{m_\pi^2}{f_\pi}\,+\,(g_8+g_{27})\,\frac{p^2_{\pi^0}}{f_\pi}   \,,\\
\mathcal{A}\big(K^+\to\pi^+\eta_{ud}^{\;*}\big)~&=~(2g_8+3g_{27})\,\frac{m_K^2}{f_\pi}\,-\,(g_8+2g_{27})\,\frac{m_\pi^2}{f_\pi}\,-\,(g_8+g_{27})\,\frac{p^2_{\eta_{ud}}}{f_\pi}   \,,\\
\mathcal{A}\big(K^+\to\pi^+\eta_{s}^{\,*}\big)~&=~\sqrt{2}\,g_{27}\,\frac{m_K^2}{f_\pi}\,-\,\sqrt{2}\,(g_8+2g_{27})\,\frac{m_\pi^2}{f_\pi}\,+\,\sqrt{2}\,(g_8+g_{27})\,\frac{p^2_{\eta_{s}}}{f_\pi}   \,.
\\
&\nonumber
\end{alignat}
\end{subequations}
\end{widetext}

The axionic decay $K^+\to\pi^+a$ is then induced by these amplitudes via axion-meson mixing:
%
\beqnarray\label{AmpKp}
\mathcal{A}(K^+\to\pi^+a)~=~~&&\theta_{a\pi}\,\mathcal{A}(K^+\to\pi^+{\pi^{0}}^{*})\Big|_{p^2_{\pi^0}=m_a^2}\nonumber\\
+\,&&\thetaUD\mathcal{A}(K^+\to\pi^+\eta_{ud}^{\;*})\Big|_{p^2_{\eta_{ud}}=m_a^2}\nonumber\\
+\,&&\thetaS\mathcal{A}(K^+\to\pi^+\eta_{s}^{\,*})\Big|_{p^2_{\eta_{s}}=m_a^2}\,.
\eeqnarray

Note that (\ref{AmpKp}) depends on $g_8$ but not on $g_8^\prime$. This implies that $\mathcal{A}(K^+\to\pi^+a)$ is only octet-enhanced in the scenario with $g_8\gg g_8^\prime$, {\it i.e.}, in the standard realization of octet enhancement in $\chiPT$ via $O_8$.  In this case, using (\ref{KsToPiPi}) and taking $g_8^\prime\to 0$ for simplicity, we can approximate (\ref{AmpKp}) as:
\beqnarray\label{AmpRatioKpOct}
&&\!\!\!\!\!\!\big|\mathcal{A}(K^+\to\pi^+a)\big|^{2\,}\bigg|_{\text{octet enh.}}\!\!\approx~\frac{1}{K_{\pi\pi}}\;\big|\mathcal{A}(K^0_S\to\pi^+\pi^-)\big|^2\nonumber\\
&&~~~~~~~~\times~\frac{~\left| 2\,g_8\,\thetaUD+\sqrt{2}\,\thetaS\Big(g_{27}-g_8\frac{m_\pi^2}{m_K^2}\Big)\right|^2}{|2\,(g_8+g_{27})|^2}\,,
\eeqnarray
where $K_{\pi\pi}\sim3$ corrects for the fact that strong $s$-wave $\pi\pi$ final state interaction, present in $K_S^0\to\pi^+\pi^-$, is absent in $K^+\to\pi^+a$ \cite{Antoniadis:1981zw}.
\\

With (\ref{AmpRatioKpOct}) and (\ref{g27}) we finally obtain:

\beqnarray\label{BrKpOct}
&&\!\!\!\Br(K^+\to\pi^+a)\Big|_{\text{octet enh.}}\nonumber\\
&&\approx\frac{\!\!\!|\mathcal{A}(K^+\to\pi^+a)|^2}{~|\mathcal{A}(K^0_S\to\pi^+\pi^-)|^2~}\bigg|_{(\ref{AmpRatioKpOct})}\Br(K_S^0\to\pi^+\pi^-)\frac{\Gamma_{K^0_S}}{\;\Gamma_{K^+}}\;\frac{\vec{p}_a}{\vec{p}_{\pi}}\nonumber\\
&&\approx~0.9\times10^{-5}~\left|\frac{\,\thetaUD\!-\,0.032\,\thetaS\,}{5\times10^{-4}}\right|^2\,.
\eeqnarray
\\
In the scenario with $g_8\ll g_8^\prime$, $\mathcal{A}(K^+\to\pi^+a)$ is not octet-enhanced. Since $g_8$ is then expected to be of the same magnitude as $g_{27}$, there might be non-negligible interference between the contributions stemming from $O_8$ and $O_{27}$. For simplicity, we will ignore this effect and consider the limiting case of $g_8\to 0$, when $O_{27}$ provides the dominant contribution to axio-hadronic $K^+$ decays.
In this case, using (\ref{KpToPiPi}), (\ref{AmpKp}) can be approximated as:

\vspace{3cm}
\beqnarray\label{AmpRatioKpNoOct}
\big|\mathcal{A}(K^+\to\pi^+a)\big|^{2\,}\bigg|_{\text{27-plet}}\approx&&~\frac{1}{K_{\pi\pi}}\;\big|\mathcal{A}(K^+\to\pi^+\pi^0)\big|^2\nonumber\\
&&\times~\Big|\theta_{a\pi}+\thetaUD\!+\frac{\sqrt{2}\,}{3}\,\thetaS\Big|^2\,,\nonumber\\
&&
\eeqnarray
from which it follows that:

\beqnarray\label{BrKpNoOct}
&&\!\!\!\Br(K^+\to\pi^+a)\Big|_{\text{27-plet}}\nonumber\\
&&~~~~\approx\frac{\!\!\!|\mathcal{A}(K^+\to\pi^+a)|^2}{~|\mathcal{A}(K^+\to\pi^+\pi^0)|^2~}\bigg|_{(\ref{AmpRatioKpNoOct})}\Br(K^+\to\pi^+\pi^0)\;\frac{\vec{p}_a}{\vec{p}_{\pi}}\nonumber\\
&&~~~~\approx~2\times10^{-8}~\left|\frac{\,\theta_{a\pi}+\thetaUD\!+\frac{\sqrt{2}\,}{3}\,\thetaS\,}{5\times10^{-4}}\right|^2\,.
\eeqnarray
\\
\\

In Figs.\,\ref{8Be4He} and \ref{KaonPlots}, we show the excluded parameter space for the axion's isoscalar mixing angles---assuming a conservative experimental bound of $\Br(K^+\to\pi^+a)\lsim10^{-5}$ (see discussion in \cite{Alves:2017avw})---for the two assumed scenarios of octet enhancement in $\chiPT$ which resulted in (\ref{BrKpOct}) and (\ref{BrKpNoOct}).

We also show in Fig.\,\ref{KaonPlots} the hypothetical reach of $K^+\to\pi^+a$ to the axion isoscalar mixing angles, assuming an experimental sensitivity\footnote{This choice of branching ratio sensitivity benchmark of $10^{-8}$ is intended to facilitate comparison between different axionic Kaon decay modes, and is not informed by any experimental sensitivity projections.} to branching ratios $\Br(K^+\to\pi^+a)\gsim 10^{-8}$. The $K^+\to\pi^+a$ contours in Fig.\,\ref{KaonPlots} make evident that this axionic Kaon decay channel is one of the most sensitive in probing the \emph{piophobic} QCD axion, and that updating the three-decades-old bounds on $K^+\to\pi^+ e^+e^-$ with $m_{e^+e^-} \lsim 50$\,MeV \cite{Yamazaki:1984qx,Yamazaki:1984vg,Baker:1987gp} could cover presently unexplored and well motivated parameter space of light BSM sectors.

\vspace{-0.2cm}
\subsection{$K_S^0$ decays}
\label{secKSdecays}

The axio-hadronic decays of the CP-even neutral Kaon can be estimated via an analogous prescription as the one used in Subsec.\,\ref{secKpPia}. First, we obtain the contributions to $\mathcal{A}(K^0_S\to\pi^0\varphi^{*})$, $\varphi=\pi^{0},\,\eta_{ud},\,\eta_{s}$, from operators (\ref{O8}), (\ref{O8prime}), and (\ref{O27}). With $K^0_S$ and $\pi^0$ on shell, we have:
\begin{widetext}
\begin{subequations}
\begin{alignat}{1}
\mathcal{A}\big(K^0_S\to\pi^0\,{\pi^{0}}^{*}\big)~&=~(g_8+g_8^\prime-2g_{27})\bigg(\frac{2\,m_K^2}{f_\pi}-\frac{m_\pi^2}{f_\pi}-\frac{p^2_{{\pi^{0}}^{*}}}{f_\pi}\!\bigg)-g_{8\,}^\prime\bigg(\frac{m_\pi^2}{f_\pi}+\frac{p^2_{{\pi^{0}}^{*}}}{f_\pi}\!\bigg)   \,,\label{KsPi}\\
\mathcal{A}\big(K^0_S\to\pi^0\,\eta_{ud}^{\;*}\big)~&=\;-\,2\,g_8\,\frac{m_K^2}{f_\pi}\,+\,(g_8+2g_{27})\,\frac{m_\pi^2}{f_\pi}\,+\,(g_8-2g_{27})\,\frac{p^2_{\eta_{ud}}}{f_\pi}   \,,\label{KsEtaUD}\\
\mathcal{A}\big(K^0_S\to\pi^0\,\eta_{s}^{\,*}\big)~&=\;-\,4\,\sqrt{2}\,g_{27}\,\frac{m_K^2}{f_\pi}\,+\,\sqrt{2}\,(g_8+2g_{27})\,\frac{m_\pi^2}{f_\pi}\,-\,\sqrt{2}\,(g_8-2g_{27})\,\frac{p^2_{\eta_{s}}}{f_\pi}   \,.\label{KsEtaS}
\end{alignat}
\end{subequations}
\end{widetext}
The axionic decay $K^0_S\to\pi^0 \,a$ is then induced by these amplitudes via axion-meson mixing:
\beqnarray\label{AmpKs}
\mathcal{A}(K^0_S\to\pi^0\,a)~=~~&&\theta_{a\pi}\,\mathcal{A}(K^0_S\to\pi^0{\pi^{0}}^{*})\Big|_{p^2_{{\pi^{0}}^{*}}=m_a^2}\nonumber\\
+\,&&\thetaUD\mathcal{A}(K^0_S\to\pi^0\eta_{ud}^{\;*})\Big|_{p^2_{\eta_{ud}}=m_a^2}\nonumber\\
+\,&&\thetaS\mathcal{A}(K^0_S\to\pi^0\eta_{s}^{\,*})\Big|_{p^2_{\eta_{s}}=m_a^2}\,.
\eeqnarray

Note that the only occurrence of $g_8^\prime$ in (\ref{AmpKs}) stems from the axion-pion mixing contribution in (\ref{KsPi}), and, as such, it is suppressed by the small $\theta_{a\pi}$ mixing angle. Because of this, $\mathcal{A}(K^0_S\to\pi^0\,a)$ parallels the behavior of $\mathcal{A}(K^+\to\pi^+a)$ of only being octet-enhanced in the scenario with $g_8\gg g_8^\prime$. In this case, using (\ref{KpToPiPi}), we can approximate (\ref{AmpKs}) as:
\beqnarray\label{AmpRatioKsOct}
\big|\mathcal{A}(K^0_S\to\pi^0\,a)\big|^{2\,}\bigg|_{\text{octet enh.}}&&\nonumber\\
~\approx~\frac{1}{K_{\pi\pi}}\;\big|\mathcal{A}(K^0_S&&\to\pi^+\pi^-)\big|^2\;\big|\thetaUD\big|^2\,,~~~~
\eeqnarray
to then obtain:
\beqnarray\label{BrKsOct}
&&\!\!\!\Br(K^0_S\to\pi^0\,a)\Big|_{\text{octet enh.}}\nonumber\\
&&~~\approx~\frac{\!\!\!\!|\mathcal{A}(K^0_S\to\pi^0\,a)|^2}{~|\mathcal{A}(K^0_S\to\pi^+\pi^-)|^2~}\bigg|_{(\ref{AmpRatioKsOct})}\Br(K_S^0\to\pi^+\pi^-)\;\frac{\vec{p}_a}{\vec{p}_{\pi}}\nonumber\\
&&~~\approx~6\times10^{-8}~\left|\frac{\,\thetaUD\,}{5\times10^{-4}}\right|^2\,.
\eeqnarray

In the alternative scenario with $g_8^\prime\gg g_8$, when $g_8$ and $g_{27}$ are expected to have comparable magnitudes, there will be non-negligible interference between the $O_8$ and $O_{27}$ contributions to the amplitudes (\ref{KsEtaUD}) and (\ref{KsEtaS}). For simplicity, we again consider the limiting case of $g_8\to 0$ to arrive at the following approximation:
\beqnarray\label{AmpRatioKsNoOct}
\!\!\big|\mathcal{A}(K^0_S\to\pi^0\,a)\big|^{2\,}\bigg|_{\text{27-plet}}&&\approx~\frac{1}{K_{\pi\pi}}\;\big|\mathcal{A}(K^0_S\to\pi^+\pi^-)\big|^2\nonumber\\
&&\times~\frac{~\left|2\,g_8^\prime\,\theta_{a\pi}-4\,\sqrt{2}\,g_{27}\,\thetaS\right|^2}{|2\,(g_8^\prime+g_{27})|^2}\,,\nonumber\\
&&
\eeqnarray
from which it follows that:

\beqnarray\label{BrKsNoOct}
&&\!\!\!\Br(K^0_S\to\pi^0\,a)\Big|_{\text{27-plet}}\nonumber\\
&&~~~~\approx\frac{\!\!\!\!|\mathcal{A}(K^0_S\to\pi^0\,a)|^2}{~|\mathcal{A}(K^0_S\to\pi^+\pi^-)|^2~}\bigg|_{(\ref{AmpRatioKsNoOct})}\Br(K^0_S\to\pi^+\pi^-)\;\frac{\vec{p}_a}{\vec{p}_{\pi}}\nonumber\\
&&~~~~\approx~0.8\times10^{-8}~\left|\frac{\,\thetaS-11\,\theta_{a\pi}\,}{2\times10^{-3}}\right|^2\,.
\eeqnarray

Note that in (\ref{BrKsNoOct}) the contribution from axion-pion mixing is non-negligible despite the suppression of $\theta_{a\pi}$ relative to $\thetaS$. As alluded to earlier, this is because (\ref{KsPi}) is the only octet-enhanced amplitude contributing to $K^0_S\to\pi^{0\,}a$ in the scenario with $g_8^\prime\gg g_8$.
\\

Present bounds on $K^0_S\to\pi^{0\,}(a\to e^+e^-)$ are difficult to infer from published experimental analyses of this final state. The observation of the SM process $K^0_S\to\pi^{0\,} e^+e^-$ by the NA48/1 experiment at the CERN SPS \cite{Batley:2003mu}, with a measured branching ratio of $\Br(K^0_S\to\pi^{0\,} e^+e^-)=\big(5.8^{\,+2.9}_{\,-2.4}\big)\times10^{-9}$, rejected events with $m_{e^+e^-}<165\,$MeV. In prior searches for this decay mode, the published analyses by NA31 \cite{Barr:1993te} and NA48 \cite{Lai:2001jf} showed the observed $m_{e^+e^-}$ distributions down to $m_{e^+e^-}\sim 0$, even though events with $m_{e^+e^-}<140\,$MeV (NA31), $m_{e^+e^-}<165\,$MeV (NA48) were rejected when extracting upper bounds on $\Br(K^0_S\to\pi^{0\,} e^+e^-)$. In particular, the $m_{e^+e^-}$ distribution in Fig.\,3 of \cite{Lai:2001jf} shows 2 events within the window of $10\;$MeV$\,<m_{e^+e^-}< 25\;$MeV, which could in principle be compatible with an axionic signal from $K_S^0$ decays with a branching ratio of $\Br(K^0_S\to\pi^{0\,}a)\sim(2-3)\times10^{-7}$. More conservatively, one could instead infer an upper bound of $\Br(K^0_S\to\pi^{0\,}a)\lsim0.8\times10^{-6}$. For the scenario with $g_8\gg g_8^\prime$, this would then translate into an upper bound on the axion isoscalar mixing angles of $\thetaUD\lsim 2\times10^{-3}$, which is $\sim3.6$ times weaker than the bound on $\thetaUD$ from $K^+\to\pi^+ a$. In the alternative scenario with $g_8^\prime\gg g_8$, and considering the limiting case of $g_8\to 0$ (cf. (\ref{BrKsNoOct})), this would then result in an upper bound on the axion isoscalar mixing angles of $\thetaS\lsim2\times 10^{-2}$, which is comparable with the bound on $\thetaS$ from $K^+\to\pi^+ a$. 
\\

However, without a proper reinterpretation of the data in \cite{Lai:2001jf}  by the NA48 Collaboration itself, we cannot have confidence that these inferred limits are accurate; therefore, we refrain from displaying them in Fig.\,\ref{KaonPlots} . Instead, in Fig.\,\ref{KaonPlots} we display the hypothetical reach of $K^0_S\to\pi^0\,a$ to the axion isoscalar mixing angles, assuming an experimental sensitivity\footnote{This choice of branching ratio sensitivity benchmark of $10^{-8}$ is intended to facilitate comparison between different axionic Kaon decay modes, and is not informed by any experimental sensitivity projections.} to branching ratios $\Br(K^0_S\to\pi^{0\,}a)\gsim 10^{-8}$, under both octet enhancement scenarios in $\chiPT$. Since there is a non-negligible contribution from $\theta_{a\pi}$ in the scenario with $g_8^\prime\gg g_8$, we have chosen the relative sign between $\theta_{a\pi}$ and $\thetaS$ that yields the most conservative reach in the parameter space of Fig.\,\ref{KaonPlots}  (cf. (\ref{BrKsNoOct})).

\subsection{$K_L^0$ decays}
\label{secKLdecays}

\subsubsection{CP-violating axio-hadronic $K_L^0$ decays}
\label{secKLdecaysCP}

Direct CP violation in the neutral Kaon system causes $K_L^0$ to inherit the axio-hadronic decay modes of $K_S^0$. The resulting branching ratios can be trivially obtained by accounting for $K_L^0-K^0_S$ mixing, parametrized by the parameter $\epsilon_K\simeq 2.23\times 10^{-3}$:
\beqnarray
\!\!\!\!\Br(K_L^0\to\pi^0\,a)~&=&~\epsilon_K^2\;\frac{\Gamma_{K_S}}{\Gamma_{K_L}}~\Br(K_S^0\to\pi^0\,a)\,,\nonumber\\
&\approx&~2.8\times10^{-3}~\Br(K_S^0\to\pi^0\,a)\,.~~~
\eeqnarray

In particular, for the two octet enhancement scenarios in $\chiPT$ considered in Subsec.\,\ref{secKSdecays}, we have:
\beqnarray
\Br(K^0_L\to\pi^0\,a)\Big|_{\text{octet enh.}}\!\!\approx\,2\times10^{-10}\left|\frac{\,\thetaUD\,}{5\times10^{-4}}\right|^2,~~~~~~~~~\label{BrKLCPoct}
\eeqnarray
and
\beqnarray
\Br(K^0_L\to\pi^0\,a)\Big|_{\text{27-plet}}\!\!\approx\,2\times10^{-11}\left|\frac{\,\thetaS-11\,\theta_{a\pi}\,}{2\times10^{-3}}\right|^2. ~~~~~~\label{BrKLCPnoOct}
\eeqnarray

An upper bound on $K_L^0\to\pi^0\,a$ can be inferred from a search for light higgs bosons in the final state $K_L^0\to\pi^0(h\to e^+e^-)$ performed by CERN's SPS NA31 experiment in 1990 \cite{Barr:1989pv}. Assuming that this analysis' efficiency to promptly decaying axions is comparable to that of much longer lived higgses of mass $\sim 17\;$MeV, the bound on the axionic decay would be $\Br(K^0_L\to\pi^0\,(a\to e^+e^-))\lsim (1-2)\times 10^{-8}$. The resulting constraints on the axion isoscalar mixing angles of $\thetaUD\lsim(3.5-5)\times 10^{-3}$ (for $g_8^\prime\to 0$) and $\thetaS\lsim(4.5-6.4)\times 10^{-2}$ (for $g_8\to 0$) are not competitive with limits from $K^+\to\pi^+ a$ (see Fig.\,\ref{KaonPlots}).

\subsubsection{CP-conserving axio-hadronic $K_L^0$ decays}
\label{secKLdecays3body}
The CP-conserving axio-hadronic decays of the CP-odd neutral Kaon can be estimated via an analogous prescription as the one used in Subsecs.\,\ref{secKpPia} and \ref{secKSdecays}. For specificity, we consider the final state with charged pions, and obtain the contributions to $\mathcal{A}(K^0_L\to\pi^+\pi^-\varphi^{*})$, $\varphi=\pi^{0},\,\eta_{ud},\,\eta_{s}$, from operators (\ref{O8}), (\ref{O8prime}), and (\ref{O27}). Putting $K^0_L$\,, $\pi^+$, and $\pi^-$ on shell, we have:
\begin{widetext}
\vspace{-0.2cm}
\begin{subequations}
\begin{alignat}{1}
\!\!\!\!\!\!\mathcal{A}\big(K^0_L\to\pi^+\pi^-{\pi^{0}}^{*}\big)~=&~~\frac{(g_8+g_8^\prime+2g_{27})}{3}\,\frac{m_K^2}{f_\pi^2}\,-\,\frac{(g_8+4g_8^\prime+11g_{27})}{3}\,\frac{m_\pi^2}{f_\pi^2}\label{KLpi0toA}\\
&+\,\frac{(g_8+g_8^\prime+11g_{27})}{3}\,\frac{p^2_{\pi^0}}{f_\pi^2}\,+\,\Big(g_8+g_8^\prime-\frac{5}{2}g_{27}\Big)\,\frac{m_\pi^2\,Y}{f_\pi^2}\nonumber\,,\\
\!\!\!\!\!\!\mathcal{A}\big(K^0_L\to\pi^+\pi^-\eta_{ud}^{\;*}\big)~=&~-\frac{(3g_8+g_8^\prime-4g_{27})}{3}\,\frac{m_K^2}{f_\pi^2}\,+\,\frac{(2g_8+4g_8^\prime-3g_{27})}{3}\,\frac{m_\pi^2}{f_\pi^2}\label{KLetaUDtoA}\\
&+\,\frac{(g_8-g_8^\prime-g_{27})}{3}\,\frac{p^2_{\eta_{ud}}}{f_\pi^2}\,-\,\Big(g_8^\prime+\frac{1}{2}g_{27}\Big)\,\frac{m_\pi^2\,Y}{f_\pi^2}\nonumber\,,\\
\!\!\!\!\!\!\mathcal{A}\big(K^0_L\to\pi^+\pi^-\eta_{s}^{\,*}\big)~=&~~\frac{\sqrt{2}\,}{3}(g_8-g_8^\prime)\,\frac{m_K^2}{f_\pi^2}\,+\,\frac{\sqrt{2}\,}{3}(4g_8^\prime-g_{27})\,\frac{m_\pi^2}{f_\pi^2}\label{KLetaStoA}\\
&-\,\frac{\sqrt{2}\,}{3}(g_8+g_8^\prime-g_{27})\,\frac{p^2_{\eta_{s}}}{f_\pi^2}\,+\,\sqrt{2}\,\Big(g_8-g_8^\prime-\frac{3}{2}g_{27}\Big)\,\frac{m_\pi^2\,Y}{f_\pi^2}\nonumber\,,
\end{alignat}
\end{subequations}
\end{widetext}
where the Dalitz plot variable $Y$ has been defined in (\ref{DalitzY}).

Furthermore, an additional effect contributing to $\mathcal{A}(K^0_L\to\pi\,\pi\,a)$ must be taken into account, namely, kinetic mixing between $K_L^0$ and the neutral pseudoscalar mesons induced by operators $O_8$ and $O_{27}$:
\beqnarray\label{KLmesonMix}
\mathcal{L}_{\chiPT}^{\,(\Delta S =1)\,}\,\supset~&-&~2\,(g_8+2\,g_{27})\, \partial_\mu K_L^0\,\partial^\mu\pi^0\\
&+&~2\,(g_8-2\,g_{27})\, \partial_\mu K_L^0\,\big(\partial^\mu\eta_{ud}\,+\,\sqrt{2}\;\partial^\mu\eta_{s}\big)\,.\nonumber
\eeqnarray

In particular, accounting for $K_L^0-\pi^0$ mixing
is crucial in order to obtain the correct dependence of the SM amplitude $\mathcal{A}(K^0_L\to\pi^+\pi^-\pi^0)$ on $g_8$ and $g_{27}$, see (\ref{AmpKLongSM}).
Similarly, the contribution to the axionic amplitude $\mathcal{A}(K^0_L\to\pi\,\pi\,a)$ stemming from $K_L^0-\eta^{(\prime)}$ mixing becomes important in the scenario with $g_8\gg g_8^\prime$ when $|\thetaUD|$, $|\thetaS|\,\lsim\OO(10^{-4})$. It can be straightforwardly obtained by re-weighting $\mathcal{A}_{\eta^{(\prime)}\to\pi\pi a}$ in (\ref{etaAmplitude}):
\beq\label{KLmixAmpl}
\mathcal{A}(K^0_L\to{\eta^{(\prime)}}^*\!\!\to\pi\,\pi\,a)\,=\,\frac{C_{K_L^0}}{\,C_{\eta^{(\prime)}}}\,\mathcal{A}(\eta^{(\prime)}\!\to\pi\pi a)\Big|_{p_{\eta^{(\prime)}}\,\to\, p_{K_L^0}}\,,
\eeq
where
\begin{widetext}
\beqnarray\label{CKLong}
C_{K_L^0}~&\equiv&~2\,(g_8-2\,g_{27})\,\big\langle \eta_{ud}+\sqrt{2}\,\eta_s \big|\bigg[ \frac{m_{K_L^0}^2}{\;m_{\eta}^2-m_{K_L^0}^2\,}\,C_{\eta}\, \big| \eta \big\rangle\,+\,\frac{m_{K_L^0}^2}{\;m_{\eta^\prime}^2-m_{K_L^0}^2\,}\,C_{\eta^\prime}\, \big| \eta^\prime \big\rangle\bigg]\nonumber\\
&\approx&~0.02\,(g_8-2\,g_{27})\,,
\eeqnarray
\end{widetext}
and we have used (\ref{etaparamet}) and (\ref{Ceta}) in obtaining the approximate equality in (\ref{CKLong}).

Unfortunately, the amplitude in (\ref{KLmixAmpl}) is subject to the same destructive interference effects, and therefore the same large uncertainties, as $\mathcal{A}(\eta^{(\prime)}\to\pi\,\pi\,a)$ estimated in Subsec.\,\ref{secEtaHad}. In particular, for the region of parameter space where (\ref{KLmixAmpl}) becomes important, these uncertainties make the estimation of $\Br(K^0_L\to\pi\,\pi\,a)$ unreliable. Nonetheless, for the sake of illustration, we will include the effects of $K_L^0-\eta^{(\prime)}$ mixing in our estimations below by benchmarking the $\RchiT$ parameters entering in (\ref{KLmixAmpl}) to: $m_{a_0}=m_{f_0}=980$\,MeV, $\Gamma_{a_0}=50$\,MeV, $\Gamma_{f_0}=100$\,MeV, and $\widehat{c}_d=\widehat{c}_m=1$.

The total amplitude $\mathcal{A}(K^0_L\to\pi^+\pi^-a)$ is then given by the sum of (\ref{KLmixAmpl}) and (\ref{KLpi0toA}), (\ref{KLetaUDtoA}), and (\ref{KLetaStoA}) reweighted by axion-meson mixing angles:
\beqnarray\label{AmpKL}
\mathcal{A}(K^0_L\to\pi^+\pi^-a)~=~&&\mathcal{A}(K^0_L\to{\eta^{(\prime)}}^*\!\to\pi\,\pi\,a)\\
+&&\;\theta_{a\pi}\,\mathcal{A}(K^0_L\to\pi^+\pi^-{\pi^{0}}^{*})\Big|_{p^2_{\pi^{0}}=m_a^2}\nonumber\\
+&&\;\thetaUD\mathcal{A}(K^0_L\to\pi^+\pi^-\eta_{ud}^{\;*})\Big|_{p^2_{\eta_{ud}}=m_a^2}\nonumber\\
+&&\;\thetaS\mathcal{A}(K^0_L\to\pi^+\pi^-\eta_{s}^{\,*})\Big|_{p^2_{\eta_{s}}=m_a^2}\,.\nonumber
\eeqnarray

Note that $\mathcal{A}(K^0_L\to\pi^+\pi^-a)$ is octet-enhanced in both scenarios, and therefore we can neglect the contributions from $g_8^\prime$ and $g_{27}$ when $g_8\gg g_8^\prime$, and likewise neglect the contributions from $g_8$ and $g_{27}$ when $g_8^\prime\gg g_8$. Despite this simplification, obtaining the dependence of $\Br(K^0_L\to\pi^+\pi^-a)$ on the axion-meson mixing angles still involves nontrivial integration of the differential decay width over the three-body final state phase space. We performed this integration numerically for both octet enhancement scenarios under the assumptions stated above, and using (\ref{AmpKLongSM}) for normalization, obtained:
\beqnarray\label{BrKLg8}
&&\Br(K^0_L\to\pi^+\pi^-a)\Big|_{g_8\,\gg\,g_8^\prime}\;\approx~ 3.54\times10^{-8}~~~\\
&&~~+~10^{-4}\times(3.83\;\thetaPI-\,7.42\;\thetaUD+\,5.715\;\thetaS)\nonumber\\
&&~~+~1.18\;\thetaPI^2\,+\,3.92\;\thetaUD^2+\,2.60\;\thetaS^2\nonumber\\
&&~~+~\thetaPI\,\big(\!\!-4.125\;\thetaUD+\,3.51\;\thetaS\big)\,-\,6.14\;\thetaUD\thetaS\,,\nonumber
\eeqnarray
and
\beqnarray\label{BrKLg8p}
\Br(K^0_L\to&&\;\pi^+\pi^-a)\Big|_{g_8^\prime\,\gg\,g_8}\nonumber\\
&&\approx~1.49\;\big(\thetaPI\,+\,\thetaUD+\,\sqrt{2}\,\thetaS\big)^2\,.
\eeqnarray

We remark that the amplitude $\mathcal{A}(K^0_L\to\pi^0\pi^0\,a)$ is simply related to $\mathcal{A}(K^0_L\to\pi^+\pi^-a)$ by isospin symmetry, which results in:
\beq
\Br(K^0_L\to\pi^0\pi^0\,a)~=~\frac{1}{2}\;\Br(K^0_L\to\pi^+\pi^-a)\,.
\eeq

In Fig.\,\ref{KaonPlots}, we show the hypothetical reach of $K^0_L\to\pi^+\pi^-a$ to the axion isoscalar mixing angles assuming an experimental sensitivity\footnote{This choice of branching ratio sensitivity benchmark of $10^{-8}$ is intended to facilitate comparison between different axionic Kaon decay modes, and is not informed by any experimental sensitivity projections.} to branching ratios $\Br(K^0_L\to\pi^+\pi^-a)\gsim 10^{-8}$, under both octet enhancement scenarios in $\chiPT$.
Since the contribution from $\theta_{a\pi}$ is non-negligible for the chosen branching ratio sensitivity benchmark of $10^{-8}$, the contours in Fig.\,\ref{KaonPlots} implicitly assume the relative sign between $\thetaPI$ and $\thetaUD$ that yields the most conservative reach. 


It is worth mentioning an exception to our estimations of $\mathcal{A}(K^0_L\to\pi^+\pi^-a)$ presented in this section. Besides the two limiting cases we have been considering of $g_8\gg g_8^\prime$ and $g_8\ll g_8^\prime$, there is also the possibility that $g_8$ and $g_8^\prime$ have comparable magnitudes. In this case, there would be non-negligible interference between the amplitudes for $K^0_L\to\pi^+\pi^-a$ originating from operators $O_8$ and $O_8^\prime$, which could substantially modify the dependence of $\Br(K^0_L\to\pi^+\pi^-a)$ on the axion-meson mixing angles.

Despite the assumptions, simplifications, and uncertainties of our estimations, the $K^0_L\to\pi^+\pi^-a$ contours in Fig.\,\ref{KaonPlots} offer a compelling motivation for upcoming Kaon experiments to search for an $m_{e^+e^-}\sim\,17$\,MeV resonance in $K^0_L\to\pi\,\pi\, e^+e^-$ final states\footnote{Dedicated reanalyses of existing data in $K^0_L\to\pi\,\pi\,e^+e^-$ final states could also be potentially sensitive to the axionic signal. See, {\it e.g}., \cite{Adams:1998eu,Takeuchi:1998vr,Lai:2003ad}.}. If these searches could achieve sensitivities down to branching ratios of $\OO(10^{-8})$, they could almost fully exclude the parameter space favored by (or verify the QCD axion explanation of) the $\Be$, $\He$, and KTeV anomalies.

\subsubsection{Di-electronic $K_L^0$ decays}
\label{secKLdecaysEE}

The last rare Kaon decay we shall consider is $K_L^0\to e^+e^-$, whose amplitude can receive a potentially non-negligible contribution from $K_L^0-\eta^{(\prime)}-a$ mixing. Indeed, using (\ref{KLmesonMix}) and momentarily neglecting the SM contribution to the amplitude, it follows that the $K_L^0\to e^+e^-$ rate induced by $K_L^0-\eta^{(\prime)}-a$ mixing would be:
\beqnarray\label{KLeeAx}
&&\Br(K_L^0\to e^+e^-)\Big|_{\mathcal{A}_\text{SM}\to\,0}\\
&&~\;\simeq\,\frac{1}{\Gamma_{K_L^0}}\,\frac{m_{K_L^0}}{8\,\pi}\;\bigg|\frac{\qe\,m_e}{f_a}\,(2\,g_8-4\,g_{27})(\thetaUD\!+\sqrt{2}\,\thetaS) \bigg|^2\nonumber\\
&&~\;\simeq\,0.9\times 10^{-11}\,\bigg|\frac{g_8-2\,g_{27}}{g_8+g_8^\prime}\bigg|^2\,(\qe)^2\,\bigg(\frac{\thetaUD\!+\sqrt{2}\,\thetaS}{10^{-3}}\bigg)^2\,. \nonumber
\eeqnarray
The estimate above should be contrasted with the observed $K_L^0\to e^+e^-$ rate \cite{Ambrose:1998cc}, as well as the range of SM predictions \cite{Sehgal:1972db,Valencia:1997xe,GomezDumm:1998gw}:
\beqnarray
\Br(K_L^0\to e^+e^-)\Big|_{\text{exp}}\;&=&~\big(0.87^{\,+\,0.57\,}_{\,-\,0.41\,}\big)\times 10^{-11}\,,\label{KLeeExp}\\
\Br(K_L^0\to e^+e^-)\Big|_{\text{SM}}\;&\sim&~(0.3\;-\;0.9)\times 10^{-11}\,.\label{KLeeSM}
\eeqnarray

From (\ref{KLeeAx}) and (\ref{KLeeExp}), we can extract a nontrivial upper bound on the axion isoscalar mixing angles in the octet enhancement scenario with $g_8\gg g_8^\prime$ (shown in Fig.\,\ref{KaonPlots}):
\beq
\big|\thetaUD\!+\sqrt{2}\,\thetaS\big|\,\Big|_{g_8\gg g_8^\prime}~\lsim~0.5\times 10^{-2}\;\bigg(\frac{1/3}{|\qe|}\bigg)\,.
\eeq

For $|\,\qe\,(\thetaUD\!+\sqrt{2}\,\thetaS)|\lsim 10^{-3}$, however, the SM contribution to the amplitude cannot be neglected; in this case, the estimation in  (\ref{KLeeAx}) is not accurate. Indeed, for a significant part of the parameter space favored by the $\Be$ and $\He$ anomalies, the axionic contribution to $K_L^0\to e^+e^-$ is subdominant to that of the SM, and in fact it is negligible in the octet enhancement scenario with $g_8\ll g_8^\prime$. A tantalizing possibility remains, however, that once measurements and theoretical predictions are improved over (\ref{KLeeExp}) and (\ref{KLeeSM}), a \emph{piophobic} QCD axion signal could appear in this channel as an excess over the SM expectation.

\section{Summary and Discussion}
\label{conclusion}

The PQ mechanism was conceived to address a problem intrinsic to the nonperturbative dynamics of QCD. Yet, presently, the prevalent view is that PQ symmetry breaking should take place at scales $f_{\text{PQ}}\,\gsim\, 10^{10}\,\Lambda_{\text{QCD}}$. Why should these two scales be so widely separated? PQ cancellation of the strong CP phase would be much more robust against spoiling effects if $f_{\text{PQ}}\,\sim\,\Lambda_{\text{QCD}}$. This possibility has long been dismissed due to stringent laboratory constraints on the visible QCD axion, in particular on its isovector couplings. More recently, however, we showed in \cite{Alves:2017avw} that the $\OO(10)$\,MeV mass range for the QCD axion remains compatible with all existing experimental constraints if the QCD axion (i) couples dominantly to the first generation of SM fermions; (ii) is short-lived, decaying with lifetimes $\lsim 10^{-13}$\,s to $e^+e^-$; and (iii) is \emph{piophobic}, {\it i.e.}, has suppressed isovector couplings due to an accidental cancelation of its mixing with the neutral pion, $\thetaPI\lsim 10^{-4}$. These conditions require nontrivial UV completions, but so does \emph{any} viable QCD axion model, whether ``heavy'' and short-lived or ultralight and cosmologically long-lived.

While this possibility forgoes the attractive feature of explaining the particle nature of dark matter, it offers a single, consistent explanation for a few persistent experimental anomalies: the observed rate for $\pi^0\to e^+e^-$, and the ``bump-like'' excesses in the $e^+ e^-$ spectra of (predominantly) isoscalar magnetic transitions of excited $\Be$ and $\He$ nuclei. Unsurprisingly, such signals have long been predicted as smoking gun signatures of the QCD axion.

In this article we estimated the axionic emission rates of the relevant $\Be$ and $\He$ transitions, taking nuclear and $\chiPT$ uncertainties into account, and showed that the \emph{piophobic} QCD axion provides a natural and compelling explanation of the observed data for these two nuclei with quite distinct properties. We also considered in detail potential axionic signals in rare decays of the $\eta$, $\eta^\prime$, $K^\pm$, and $K^0_{S,L}$ mesons. The (often ignored) hadronic and $\chiPT$ uncertainties involved in estimations of these rare meson decays impede accurate predictions of axio-hadronic signals; nonetheless, the ranges we have obtained for several of the processes investigated can be probed in the near future in a variety of experimental programs, including $\eta/\eta^\prime$ and Kaon factories.

\begin{acknowledgments}
I am grateful to Anna Hayes and Gerry Hale for explanations of electromagnetic transitions of $^8\text{Be}$ and $^4\text{He}$ nuclei. I also thank Maxim Pospelov for discussions on the experimental sensitivity of the COSY-WASA analysis to axionic $\eta$ decays, and Corrado Gatto for discussions about the REDTOP experiment.
This work was supported by an Early Career Research award from Los Alamos National Laboratory's LDRD program, and by the DOE Office of Science High Energy Physics under Contract No. DE-AC52-06NA25396.
\\
\vspace{0.7 cm}
\\
{\bf{Note added}}~~In earlier preprint versions of this article, as well as in its published version in PRD, an upper bound on the axion lifetime of $\tau_a\lsim 4\times 10^{-14}$\,s had been inferred based on NA64's limit on the lifetime of a visibly decaying {\it vector boson} with mass of $\sim 17$ MeV \cite{Banerjee:2019hmi,Depero:2020zfy}. After this article was accepted for publication, it came to the author's attention that the reinterpreted NA64's bounds on the lifetimes of {\it pseudoscalars} would generally be weaker than corresponding bounds on {\it vectors}, due to pseudoscalars' smaller production cross section relative to that of vectors with the same lifetime---see the \href{https://indico.cern.ch/event/1002356/contributions/4229597/attachments/2200248/3721193/gninenko-na64-pbc-020321.pdf}{preliminary NA64 results} presented at the \href{https://indico.cern.ch/event/1002356/}{2021 Physics Beyond Colliders Annual Workshop}, virtually held by CERN during March 1-4, 2021. Based on these unpublished preliminary results by the NA64 Collaboration, the upper bound on the lifetime of a  17~MeV piophobic axion appears to be closer to $\tau_a\lsim 10^{-13}$\,s. The author is grateful to Sergei Gninenko for clarifying this issue.
\end{acknowledgments}

\bibliography{REFS}

\end{document}